\documentclass[final,3p]{elsarticle}

\usepackage[hidelinks]{hyperref}
\usepackage{amsmath,amssymb,amsfonts}
\usepackage{bm}
\usepackage{graphicx, wrapfig}
\usepackage{cleveref}
\usepackage[utf8]{inputenc}
\usepackage{geometry}
\usepackage{enumerate}
\usepackage{bbm}
\usepackage{xcolor}
\usepackage{tabularx,makecell}
\usepackage[linesnumbered,ruled,vlined,algo2e]{algorithm2e}
\usepackage[toc,page]{appendix}
\usepackage{makecell}
\usepackage{ntheorem}

\theoremstyle{break}
\theoremheaderfont{\normalfont\bfseries\hspace{-\theoremindent}}
\theorembodyfont{\normalfont}
\newtheorem{encoder}{Encoder}
\newtheorem{decoder}{Decoder}

\newcolumntype{Y}{>{\centering\arraybackslash}X}
\journal{Journal of Computational Physics}
\biboptions{sort&compress}

\newcommand{\R}{\mathbb{R}}

\newcommand{\eps}{\epsilon}

\newcommand{\bb}[1]{\mathbf{#1}}

\crefname{algocf}{alg.}{algs.}
\Crefname{algocf}{Algorithm}{Algorithms}
\crefname{encoder}{Encoder}{Encoders}
\crefname{decoder}{Decoder}{Decoders}

\begin{document}	
	\begin{frontmatter}
		
		\title{Machine Learning Techniques to Construct Patched Analog Ensembles for Data Assimilation}
		
		\author{L. Minah Yang\corref{c1}}
		\ead{lucia.yang@colorado.edu}
		\cortext[c1]{Corresponding author}
		\author{Ian Grooms}
		\address{Department of Applied Mathematics, University of Colorado, Boulder, CO 80309}
		\begin{abstract}
    Using generative models from the machine learning literature to create artificial ensemble members for use within data assimilation schemes has been introduced in \cite[Grooms QJRMS, 2020]{grooms2020analog}
    as constructed analog ensemble optimal interpolation (cAnEnOI).
    Specifically, we study general and variational autoencoders for the machine learning component of this method, and combine the ideas of constructed analogs and ensemble optimal interpolation in the data assimilation piece. 
    To extend the scalability of cAnEnOI for use in data assimilation on complex dynamical models, we propose using patching schemes to divide the global spatial domain into digestible chunks.
    Using patches makes training the generative models possible and has the added benefit of being able to exploit parallelism during the generative step. 
    Testing this new algorithm on a 1D toy model, we find that larger patch sizes make it harder to train an accurate generative model (i.e. a model whose reconstruction error is small), while conversely the data assimilation performance improves at larger patch sizes.
    There is thus a sweet spot where the patch size is large enough to enable good data assimilation performance, but not so large that it becomes difficult to train an accurate generative model.
    In our tests the new patched cAnEnOI method outperforms the original (unpatched) cAnEnOI, as well as the ensemble square root filter results from \cite{grooms2020analog}.
			
		\end{abstract}
		
		\begin{keyword}
			data assimilation \sep machine learning \sep analogs
		\end{keyword}
		
	\end{frontmatter}
	
	\section{Introduction}
\label{sec:intro}
Data assimilation is widely used in geophysical problems as a reliable method to combine theoretical knowledge of the dynamical system at hand with the available real-life observations.
The main data assimilation methods are split into those that require an ensemble such as the ensemble Kalman filter (EnKF, see \cite{Evensen2003}), and variational methods such as 3D-Var and 4D-Var \cite{assimilation2010making}.
As ensemble forecasts can be prohibitively expensive, the use of time-independent ensemble perturbations to reduce costs has been introduced as ensemble optimal interpolation (EnOI, see \cite{Oke2002,Evensen2003}) and implemented in ocean data assimilation (see, e.g., \cite{Oke2010}).
Naturally, this is less accurate than using an ensemble forecast and to remedy this, it was proposed in \cite{grooms2020analog} to use analog ensembles (AnEnOI) which are time-dependent.
Analogs are model states that are close to the current forecast; in AnEnOI analogs of the forecast are found from a library of model states and used to form an analog ensemble.
The use of analogs in AnEnOI is a departure from their traditional role in forecasting, where analogs are used to produce a forecast (see \cite{delle2013probabilistic,eckel2016hybrid,zhao2016analog}).
Moreover, \cite{grooms2020analog} introduced a new method of using variational autoencoders (VAEs, a machine learning model in \cite{Kingma2014}) to construct analogs for AnEnOI, rather than finding analogs in a library.
This method is referred to as cAnEnOI and it has been shown to produce better results than both EnOI and AnEnOI when tested on a multiscale variant \cite{grooms2015framework} of the Lorenz '96 model \cite{lorenz1996predictability}. \par
In this paper, we introduce a patching scheme to the analog construction component of cAnEnOI which we call p-cAnEnOI.
As geophysical applications often require high resolution representations of complex dynamical structures, it is likely difficult for a single VAE to capture all of the significant information over the entire spatial domain.
For example, consider training a generative model to generate realistic states of an eddy resolving ocean model.
A single three-dimensional variable (e.g. temperature) from the global ocean simulation in \cite{johnson2016climatological} at nominal 0.1-degree resolution requires approximately 4 GB of memory.
There are four three-dimensional variables in a single model state (for this model; others may have more), not to mention the other two-dimensional variables like sea-surface height, and the resolution of ocean models is continually increasing.
Not only would training a generative model for something like this require massive computing powers in terms of computations during training, simply moving batches of samples around during training could cause issues in the communication sector.

To avoid the difficulty of dealing with the global model state, we simply divide the domain into equal-sized patches, and train a generative model to reproduce a single patch.
A global model state can be constructed out of patches.
Catastrophic discontinuities across patch boundaries are avoided by forcing each patch to be an analog of the corresponding patch in a single model forecast.
Some benefits of p-cAnEnOI include more robust training due to simpler complexity of the machine learning model and increase in the number of training samples.
Also, when constructing the global model state, patches can be constructed in parallel.

We test p-cAnEnOI in the same multiscale Lorenz `96 model as \cite{grooms2020analog}.
We vary the patch sizes in p-cAnEnOI and study how that affects the performance of the machine learning models with respect to reconstruction error (the ability of the VAE to exactly reconstruct the input model state), and the performance of the resulting data assimilation scheme. 
Using smaller patch sizes makes training quicker and results in VAEs with better (lower) reconstruction errors, but introduces more discontinuities across patch boundaries which can degrade the spatial correlations in the analog ensemble. \par

While cAnEnOI in \cite{grooms2020analog} used only VAEs to construct analogs, we introduce the use of general (non-variational) autoencoders in constructing analogs.
These are quicker to train due to their simpler structure but require additional steps to use as a generative model: the covariance of the training data encoded into the latent space must be computed then factorized using the Cholesky decomposition; but these extra operations are not prohibitive due to the small size of the encoded latent space. 
We find that p-cAnEnOI with general autoencoders performs just as well as the variant that uses VAEs instead.\par 

Overall, we find that p-cAnEnOI can outperform the analog methods described in \cite{grooms2020analog} as well as an ensemble square root filter, which costs 100 times more for an ensemble size of 100.
In the toy model, we found that the largest patch size (equal to 1/4 of the domain) produced the best results, and p-cAnEnOI with patch sizes of 1/8 and 1/16 still performed better than global cAnenOI but patch size 1/32 performed worse despite having the best reconstruction error.
This indicates that the patch size has an effect on the data assimilation quality where it must be large enough to preserve spatial correlations but small enough to allow for efficient training.
 
\section{Background}
\label{sec:background}
In this section, we summarize the relevant data assimilation and machine learning methods in order to sufficiently describe the new method, p-cAnEnOI, in \cref{methods:patching}. 
This new method is a slight modification of cAnEnOI, which inserts constructed analogs (typically used in forecasting) in ensemble-based data assimilation methods such as ensemble optimal interpolation (EnOI).
In addition, variational autoencoders, a machine learning model, is used to construct the analogs. 
We discuss the data assimilation methods in \cref{sec:DA}, the machine learning methods in \cref{methods:AE,methods:VAE}, cAnEnOI in \cref{methods:cAnEnOI}, and the toy model in \cref{methods:toymodel}.
\subsection{Ensemble-based Data Assimilation Methods}
\label{sec:DA}
In this section, we discuss some relevant ensemble-based data assimilation methods.
The Ensemble Kalman Filter (EnKF, see \cite{Evensen1994,houtekamer1998data,burgers1998analysis}) takes the Monte Carlo approach to approximating the background covariance matrix by forecasting an ensemble of state variables and computing its sample covariance matrix.
This corresponds to forming $\bb{B}$ in \cref{eqn:kalman_u}.
The Kalman update formula for the $i^{th}$ ensemble member is
\begin{equation}
\bb{x}_{a}^{(i)} = \bb{x}_b^{(i)} + \bb{K}(\bb{y}^{(i)}-\bb{H}\bb{x}_b^{(i)}),\qquad \bb{K}=\bb{BH}^{\top}(\bb{HBH}^{\top}+\bb{R})^{-1},\label{eqn:kalman_u}
\end{equation}
where $\bb{x}_a^{(i)}$ is the updated ensemble member, $\bb{x}_b^{(i)}$ is the forecast of the $i^{th}$ ensemble member from the previous time step, $\bb{y}^{(i)}$ is the perturbed observation where the perturbations have zero mean and covariance $\bb{R}$, and $\bb{H}$ is the observation model.
Accurately approximating the background covariance matrix in this approach may require the size of the ensemble to be large, which then increases the number of model forecasts to be made. 
The cost of the EnKF update is usually small compared to the cost of a forecast, so the total cost of using the EnKF increases linearly with respect to the size of the ensemble, $N_e$, and may quickly become impractical.
The methods we discuss in this section are motivated by reducing the computational load of forecasting $N_e$ ensemble members. 
Since we compare our results in \cref{sec:NR} to the numerical results of \cite{grooms2020analog}, the analog methods discussed in the aforementioned paper are included.
The quality of the various data assimilation methods is evaluated by the root mean squared error (RMSE) computed across $T$ assimilation cycles as shown in\cref{eqn:DA_RMSE},
\begin{equation}
\label{eqn:DA_RMSE}
RMSE = \frac{1}{T}\sum_{i=1}^T \left(\frac{1}{N_x}\sum_{j=1}^{N_x} \left(RS_j^{(i)}-AM_j^{(i)}\right)^2\right)^{1/2},
\end{equation}
where the difference between the reference simulation $RS$ and the analysis mean $AM$ is averaged for a state with $N_x$ spatial nodes and the subscript and paranthesized superscripts each refer to spatial and temporal indices.
\subsubsection{Ensemble Square Root Filter}\label{methods:ESRF}
Ensemble square root filters (reviewed in \cite{tippett2003ensemble}) modify EnKF to ensure that the posterior covariance exactly satisfies the Kalman filter update formula
\begin{equation}
    \bb{B}_a = \left(\bb{I} - \bb{KH}\right)\bb{B}_f
\end{equation}
and avoids having to build an ensemble of perturbed observations. 
In particular, \cite{grooms2020analog} implements the serial ensemble square root filter of \cite{whitaker2002ensemble}, (referred to as ESRF hereafter) to serve as a point of reference for the performance of data assimilation methods that use an ensemble forecast at every step.
Localization is added to zero out spuriously high covariances across large distances using the following localization function
\begin{equation}\label{eqn:loc}
	\ell(d_i) = \exp\left(-\frac{1}{2}\left(\frac{d_i}{L}\right)^2\right),
\end{equation}
where $d_i$ is the distance from the observation being assimilated to the $i^{th}$ state variable, and $L$ is the localization radius, a tunable parameter.
This can be efficiently applied with elementwise multiplication within ESRF.
In addition, inflation is applied to the posterior as suggested in \cite{el2019comparing} by multiplying the background covariance matrix by some $r\geq 1$. 
Since the optimal results for ESRF with an ensemble of size $N_e=200$ were not much better than that with $N_e=100$ in \cite{grooms2020analog}, we use $N_e=100$ for the remainder of this paper.
No new ESRF experiments were performed here; we only quote ESRF results from \cite{grooms2020analog}.

\subsubsection{EnOI and Analogs}\label{methods:EnOI}
Ensemble optimal interpolation (EnOI, see \cite{Oke2002,Evensen2003}) replaces the background covariance matrix in \cref{eqn:kalman_u} with the sample covariance computed from a fixed set of states drawn from a catalog of model states.
This can be interpreted as approximating the background forecast error with the covariance present in a climatological time scale and is time-independent since the same set of states is used for every assimilation cycle.
Since the climatological spread is likely larger than what is expected in the error in a single forecast step, we scale the constructed $\bb{B}$ to have a specified forecast spread $f$, which is another tunable parameter, equal to the square root of the mean along the diagonal of $\bb{B}$. \par

Analogs refer to model states that are close to the current state and they have been used in forecasting (see \cite{delle2013probabilistic,eckel2016hybrid,zhao2016analog,van2003performance}) following its introduction in \cite{lorenz1969atmospheric}.
They have primarily been used for forecasting in the context of a catalog of data that contain correlated pairs $C=\{(\bb{x}^{(i)},\bb{x}^{(i+1)})\}_{i=1,...}$.
Basic analog forecasting is performed by finding a pair in the catalog such that $\bb{x}^{(i)}$ matches the current state $\bm{x}^*$ as closely as possible, and then using the corresponding component $\bb{x}^{(i+1)}$ as the forecast.
Constructed analogs generalize this method by finding many analog pairs in $C$ whose $\bb{x}^{(i)}$ components are close to $\bb{x}^*$, and constructing the forecast with some weighted average of the corresponding $\bb{x}^{(i+1)}$ component of the pairs.
Analog forecasting is not used in AnEnOI, cAnEnOI, or p-cAnEnOI, though it has been used in data assimilation in \cite{lguensat2017analog}.\par

It was proposed in \cite{grooms2020analog} to replace the time-independent ensemble perturbations in EnOI with analog ensemble perturbations.
The resulting method (AnEnOI) makes a single forecast $\bm{x}^{(i)}$ (using the dynamical model, not an analog forecast), and then finds $N_e$ nearest-neighbors in a catalog.
The mean of the ensemble of nearest neighbors is then subtracted and replaced by the actual forecast, resulting in an ensemble whose mean is the single forecast and whose perturbations are drawn from a set of model states that are similar to the forecast.
The ensemble covariance matrix $\bb{B}$ is then scaled to have forecast spread $f$, which is a tunable parameter. 
AnEnOI exhibited approximately 15\% improvement in performance compared to EnOI in the tests of \cite{grooms2020analog}. \par

Next, we introduce autoencoders and variational encoders in \cref{methods:ML} before continuing with the descriptions of the data assimilation methods in \cref{methods:cAnEnOI}.

\subsection{Autoencoders and Variational Autoencoders}
\label{methods:ML}
While \cite{grooms2020analog} exclusively used variational autoencoders as the machine learning component of the constructed analog method, we introduce a method that uses a general autoencoder.
We briefly summarize both machine learning methods in \cref{methods:AE,methods:VAE}.

\subsubsection{General Autoencoders}\label{methods:AE}
\begin{figure}[h]
 	\centering
 	\vspace{-10pt}
 	\includegraphics[width=\textwidth]{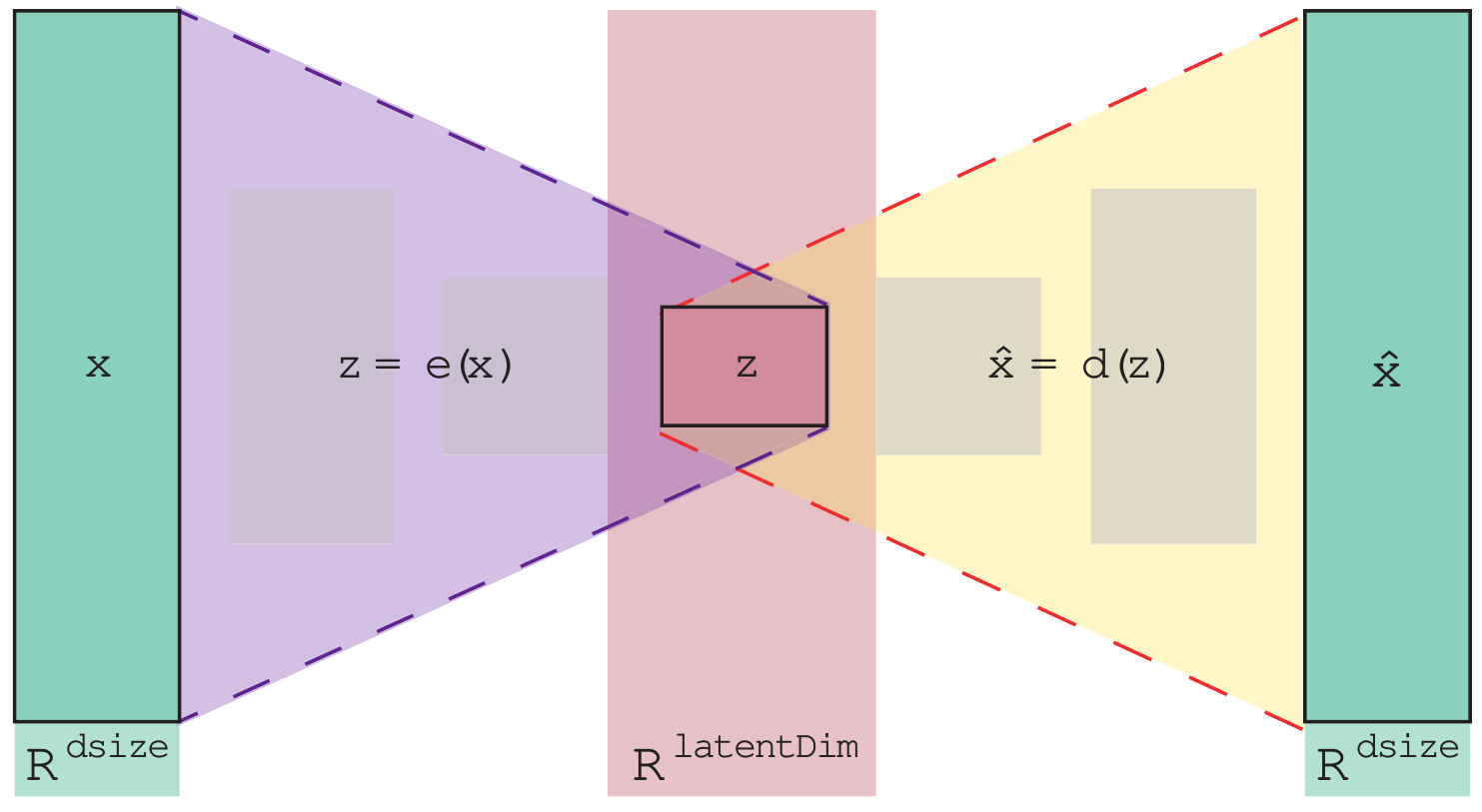}
 	\vspace{-10pt}
 	\caption{\label{fig:ae} Example architecture of a general autoencoder. Purple region represents the encoder, and the yellow region represents the decoder.}
\end{figure}

Autoencoders are a type of unsupervised learning in which deep neural networks are trained to learn an encoding of the data. 
To achieve this, the output of the neural network is set to the input, and the hidden layers of the neural network form two distinct components: the encoder and the decoder.
The encoder is comprised of (hidden) layers that code the input variables into the latent space.
The structure of the encoder is often designed to result in a latent space with fewer dimensions than that of the input space, performing a dimension reduction of a sort, though dimension reduction is not the goal here.
The decoder transforms the encoded variable back to its original input through more layers.
The encoder and decoder pair are shown as two mappings in \cref{eqn:ae_encoder,eqn:ae_decoder},
\begin{align}
e(\bb{x}) &= \bb{z},\;\;\bb{x}\in\R^{\text{{\tt dsize}}}, \bb{z}\in\R^{\text{{\tt dsize}}} \label{eqn:ae_encoder}\\
d(\bb{z}) &= \hat{\bb{x}},\;\; \hat{\bb{x}} \in \R^{\text{{\tt dsize}}}, \label{eqn:ae_decoder}
\end{align}
$\bb{x}$ is the original variable, $\bb{z}$, its code, and $\hat{\bb{x}}$, the reconstructed variable, and ${\text{\tt latentDim}} \leq {\text{\tt dsize}}$.
In practice, an autoencoder can be thought of and trained as a normal deep neural network with a special hidden layer that has lower dimension than the size of the input variable.
\Cref{fig:ae} illustrates the architecture of an autoencoder.

Autoencoders can be used for specialized lossy compression (meaning exact recovery is not guaranteed) or dimension reduction tasks for specific datasets they were trained on. 
Their use in anomaly detection and image processing has also been studied.
We use autoencoders with convolutional layers for the encoder and the decoder halves.

Many types of autoencoders have been developed for use in data science applications, i.e. sparse autoencoders for classification and variational autoencoders (VAEs) as a generative model.
We describe VAEs in the next section, and refer to autoencoders that are not VAEs as general autoencoders. 

\subsubsection{VAE}\label{methods:VAE}
A variational autoencoder is a type of an autoencoder that imposes extra conditions to regularize the latent space (see \cite{Kingma2014,Kingma2019}) to resemble a Gaussian distribution.
These extra conditions are imposed via the reparametrization trick (see \cite{Kingma2014}), described in \cref{eqn:encoder,eqn:vae_sampling} and by adding a extra loss term called the Kullback-Leibler (KL) divergence, which measures the distance from one probablistic distribution to another. 
In contrast to \cref{eqn:ae_encoder}, the encoder in a VAE outputs two variables of size {\tt latentDim}.
\Cref{eqn:encoder} shows the VAE variant of an encoder that encodes the input variable $\bb{x}$ into two variables, $\bb{z}_{\mu}$ and $\bb{z}_{\log(\bb{\Sigma})}$, which are then combined via \cref{eqn:vae_sampling} to mimic sampling from a normal distribution.
\begin{equation}
e(\bb{x})=[\bb{z}_{\mu}\;\; \bb{z}_{\log(\bb{\Sigma})}] \in \R^{{\tt latentDim} \times 2}
\label{eqn:encoder}
\end{equation}
\begin{equation}\label{eqn:vae_sampling}
\bb{z} = \bb{z}_{\mu} + \eps\circ\exp{(0.5 * \bb{z}_{\log(\bb{\Sigma})})},\;\; \eps \sim N(\bb{0},\bb{I})
\end{equation}
where $\circ$ represents an elementwise product.
(We also denote $e(\bb{x})_{\mu}:=\bb{z}_{\mu}$ and  $e(\bb{x})_{\log(\bb{\Sigma})}:=\bb{z}_{\log(\bb{\Sigma})}$.)
This encoded state, $\bb{z}$ is then fed through the decoder, which is constructed exactly as in \cref{eqn:ae_decoder}.

\begin{figure}[h]
 	\centering
 	\vspace{-10pt}
 	\includegraphics[width=\textwidth]{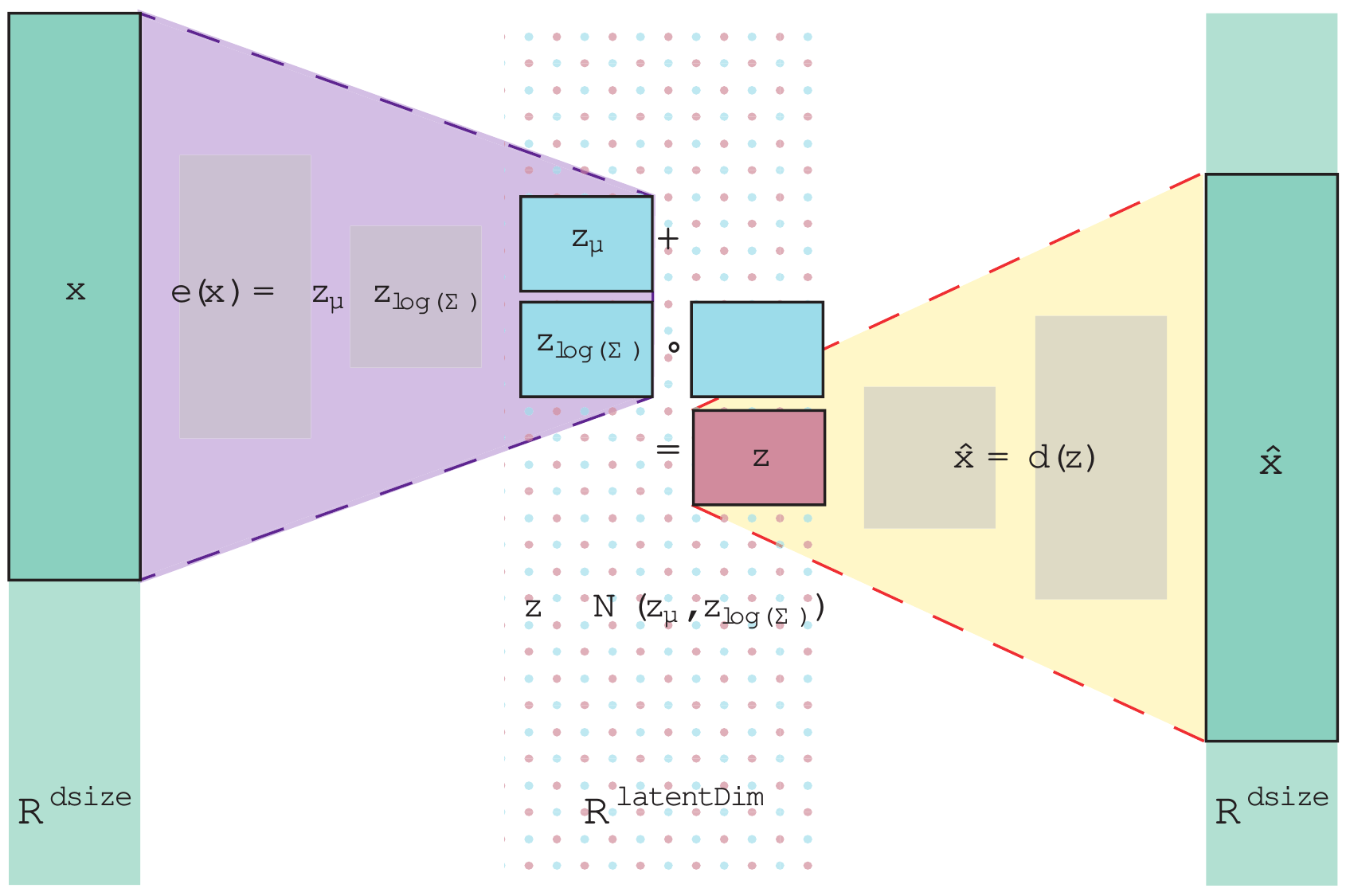}
 	\vspace{-10pt}
 	\caption{\label{fig:vae} Example architecture of a variational autoencoder (VAE).}
\end{figure}

The regularized structure of the latent space guarantees that a small perturbation added in the latent space should result in small differences once decoded. 
That is, $d(e(\bb{x})+\eps)-d(e(\bb{x}))$ should be reasonably small for a small $\eps$.
\Cref{fig:vae} provides a visualization of the structure of VAEs as a basis of comparison against \cref{fig:ae}. 
\subsection{cAnEnOI}\label{methods:cAnEnOI}

In \cite{grooms2020analog}, it is suggested that VAEs be used as a generative model for constructing analogs in the following way.
To generate $N_e$ ensemble members close to the current forecast state, $\bb{x}$, $N_e$ distinct, independent, samples of random noise with amplitude $r_z$ would be added to the mean encoded state ($e(\bb{x})_{\mu}\equiv\bb{z}_{\mu}$) in latent space and then decoded to return to the original dimension, as shown in \cref{eqn:gen_en}.
Each ensemble member is perturbed by a unique random noise, denoted by $\bb{\eps}^{(i)}$ for $i=1,\cdots,N_e$.
\begin{equation}
\bb{x}^{(i)} = d\left(e(\bb{x})_{\mu}+r_z\bb{\eps}^{(i)}\right),\;\; \bb{\eps}^{(i)}\sim N(\bb{0},\bb{I}),\; i\in\{1,\cdots,N_e\}
\label{eqn:gen_en}
\end{equation}
The sample covariance matrix of these constructed analogs approximates the background covariance matrix in \cref{eqn:kalman_u}.
The amplitude $r_z$ of the noise in latent space is a tunable parameter.
In \cref{methods:patching_ae} we propose a slightly different version of \cref{eqn:gen_en} to construct an analog ensemble using a general autoencoder.

\subsection{1D Toy Model: Multiscale Lorenz '96 }\label{methods:toymodel}
\begin{figure}[h]
 	\centering
 	\vspace{5pt}
 	\includegraphics[width=\textwidth]{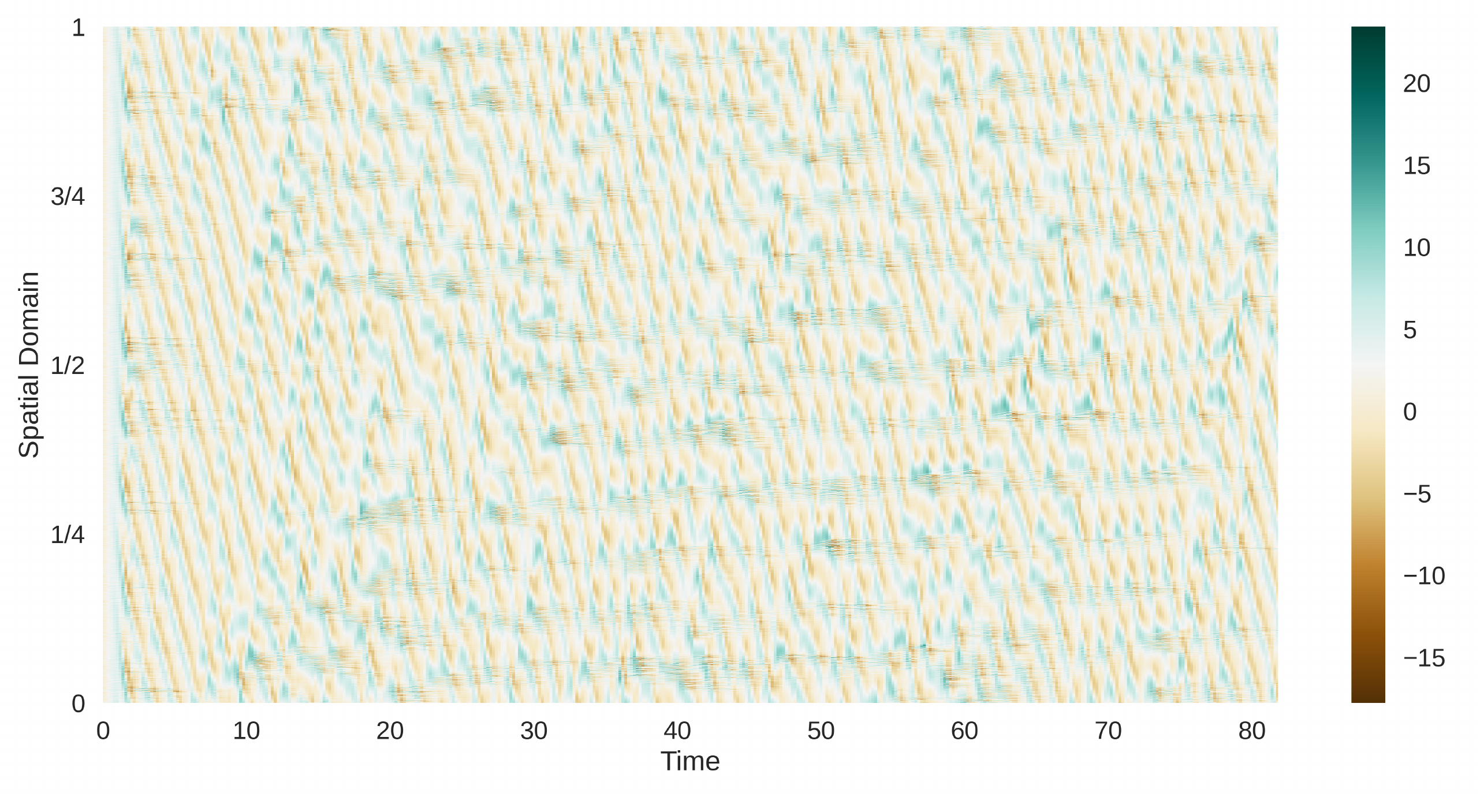}
 	\caption{\label{fig:evol} Evolution of the multiscale Lorenz-'96 model initialized by a sample from a standard normal distribution at time $T=0$ and run until final time $T=81.8$.}
\end{figure}

The multiscale Lorenz '96 model introduced in \cite{grooms2015framework} is described by a single system of ODEs in \cref{eqn:toymodel} 
\begin{equation}
    \dot{\bb{x}} =h\bb{N}_S(\bb{x})+J\bb{T}^{\top}\bb{N}_L(\bb{Tx})-\bb{x}+F\mathbbm{1},
    \label{eqn:toymodel}
\end{equation}
where $h,F\in\R$ are coupling and forcing parameters, $J\in\mathbb{N}$ is the ratio between the two scales, and $\mathbbm{1}$ is a vector of ones.
The small and large scale nonlinearities are described by \cref{eqn:N1,eqn:N2}, and $\bb{T}$ is a mapping that projects $\bb{x}$ onto the largest $K$ Fourier modes and interpolates the result to $K$ equally-spaced points.
The nonlinearities are defined by
\begin{align}
    \bb{N}_S(\bb{x})_i &= -\bb{x}_{i+1}(\bb{x}_{i+2}-\bb{x}_{i-1}),\;\;i\in\{1, \cdots, JK\},\label{eqn:N1}\\
    \bb{N}_L(\bb{X})_k &= -\bb{X}_{k-1}(\bb{X}_{k-2}-\bb{X}_{k+1}),\;\;k\in\{1, \cdots, K\}\label{eqn:N2}
\end{align}
where the indices are periodic, i.e. $\bb{x}_{i+JK}=\bb{x}_i$ and $\bb{X}_{k+K}=\bb{X}_k$.
We use this model exclusively to train the VAEs and to test our new method, patched-cAnEnOI (p-cAnEnOI), in \cref{sec:NR}.
\Cref{fig:evol} shows the evolution of the model \cref{eqn:toymodel} configured with $h=0.5$, $F=8$, $J=64$, and $K=41$, and initialized from a standard normal distribution at time 0.
The multiscale model typically displays about 8 large-scale waves in the domain, similar to the standard Lorenz `96 model, with additional small-scale instabilities appearing intermittently in time and space.

\section{Method: Patched Analogs (p-cAnEnOI)}\label{methods:patching}
Training a VAE that learns the dynamics over the entire spatial domain for use in cAnEnOI is feasible for small problems such as the multiscale Lorenz '96 from \cite{grooms2015framework} as is demonstrated in \cite{grooms2020analog}.
However, a direct implementation of cAnEnOI on more complex, high resolution models poses several problems.
First, the number of parameters being trained for convolutional neural networks grows large as the size of the input variable increases, making training computationally expensive.
Second, training a VAE to learn the complex dynamics of the global spatial domain may require extremely large libraries of training samples that may not be available.
For example, a GCM may have several dynamically significant locations that each deserve careful analysis but the global variable as a whole is too specific of a state for a VAE to be trained to learn.
Our new method, p-cAnEnOI, attempts to address these issues.
We describe the method in detail in \cref{methods:patching_d} and discuss its impact on the covariance matrix within \cref{eqn:kalman_u} in \cref{methods:cov}.

\subsection{Constructing Patched Analogs with VAEs}\label{methods:patching_d}
This new method modifies how analogs are generated at each assimilation cycle. 
Recall that to generate $N_e$ analogs, cAnEnOI uses a VAE to encode the state variable at time of assimilation, add $N_e$ distinct random noise, and decode those perturbations. 
Instead, p-cAnEnOI first partitions the state variable of size {\tt dsize} into $p$ equal sized patches of size {\tt psize}$=${\tt dsize}$/p$ given that $p$ is a divisor of {\tt dsize}.
These $p$ patches are then individually encoded with a VAE whose input and output variables correspond to {\tt psize}.
To generate $N_e$ ensemble members, $N_e$ distinct random noises are added to each encoded patch, then decoded. 
The decoded patch perturbations are rearranged to cover the original spatial domain.

Consider a trivial partitioning scheme which places patch boundaries to the left of locations $\left\{k\times \right.${\tt psize}$\left.\right\}_{k=1}^{p-1}$.
This forces discontinuities only at those $(p-1)$ locations for all constructed analogs and unevenly distributes the errors associated with these discontinuities across the domain.
To remedy this, we shift the locations of the patch boundaries. 
Since the toy model is periodic, any shift $s$ in $\{0,\cdots,\text{{\tt psize}}-1\}$ creates a unique set of $p$ patch boundaries at $\{s+k\times \text{{\tt psize}}\}_{k=0}^{p-1}$.
That is, the $p$ patches of a state $\bb{x}$ with a right shift $s$ range over the nodes as indicated in \cref{eqn:p_nodes}, where $\bb{x}^{(j,s)}$ is the $j^{th}$ patch created with shift $s$ and $\bb{x}_k$ is the $k^{th}$ element of $\bb{x}$.
\begin{equation} \label{eqn:p_nodes}
\bb{x}^{(j,s)} = 
\begin{cases}
    [\bb{x}_{s+j\times \text{{\tt psize}}+1},\cdots,\bb{x}_{s+(j+1)\times \text{{\tt psize}}}],& j = 0,\cdots, p-2 \\
    [\bb{x}_{s+(j-1)\times \text{{\tt psize}} + 1},\cdots,\bb{x}_{\text{{\tt dsize}}},\bb{x}_1,\cdots,\bb{x}_{s}], & j = p-1
\end{cases}
\end{equation}
The patching is undone by simply concatenating all $p$ patches in order and by shifting to left by $s$.
We discuss the impact of shifted partitioning on the covariance matrix in \cref{methods:cov}, patching in non-periodic domains in \cref{sec:conclusion}, and \cref{algo:p-gen} formalizes this new method of constructing analogs.
\begin{algorithm2e}
	\DontPrintSemicolon 
	\KwIn{$\bb{x}$, $N_e$, $p$, $r_z$, $S$ \hfill\textbf{Output:} $\{\bb{a}^{[i]}\}_{i=1}^{N_e}$}
 	\For{$i=1 : N_e$}{
 	\tcc*{Partition $\bb{x}$ into p patches using shift $s_i\in S$ (See \cref{eqn:p_nodes}).
 }
 	$[\bb{x}^{(1,s_i)},\cdots,\bb{x}^{(p,s_i)}] \gets {\tt patch}(\bb{x},s_i)$\\
	\For{$j=1:p$}{
	    $[z_{\mu},z_{\log(\bb{\Sigma})}]\gets e(\bb{x}^{(j,s_i)})$\\
	    $\bb{x}^{(j,s_i)} \gets d(z_{\mu} + r_z\eps^{(j,i)})$,\tcp*{$\eps^{(j,i)} \sim N(\bb{0},\bb{I})$}
	}
	$\bb{a}^{[i]} \gets$ {\tt undo\_patching}($\{\bb{x}^{(j,s_i)}\}_{j=1}^p$)
 	}
	\Return $\{\bb{a}^{[i]}\}_{i=1}^{N_e}$
 	\caption{$\{\bb{a}^{[i]}\}_{i=1}^{N_e}=${\tt gen\_patch\_ensemble}($\bb{x}$, $N_e$, $p$, $r_z$, $S$). This algorithm constructs $N_e$ analogs by using a {\tt vae} on $pN_e$ patches with shifts in $S$.}
	\label{algo:p-gen}
\end{algorithm2e}

\Cref{algo:p-gen} assumes that we have a trained VAE ready to be used on the patches. 
This adds only few extra steps to the training portion of cAnEnOI: a new library must be generated from the original library of analogs.
If $1/p$ is the ratio between the patch size we wish to use and the size of the domain, then each model state in the original training library can generate at least $p$ patches, creating a library that has at minimum $p\times$ the number of elements of the original library. 
This library can now be split into training and validation sets, and the appropriate VAE can now be trained.
Note that whatever patching scheme is used to generate this new library should be used within \cref{algo:p-gen} as well. 
Furthermore, there are many opportunities for parallelization in \cref{algo:p-gen}. 
The inner loop can be computed simultaneously with $p$ nodes, and even the outer loop can be computed in parallel if copies of model state $\bb{x}$ can be quickly distributed.

\begin{figure}[t]
    \centering
    \includegraphics[width=\textwidth]{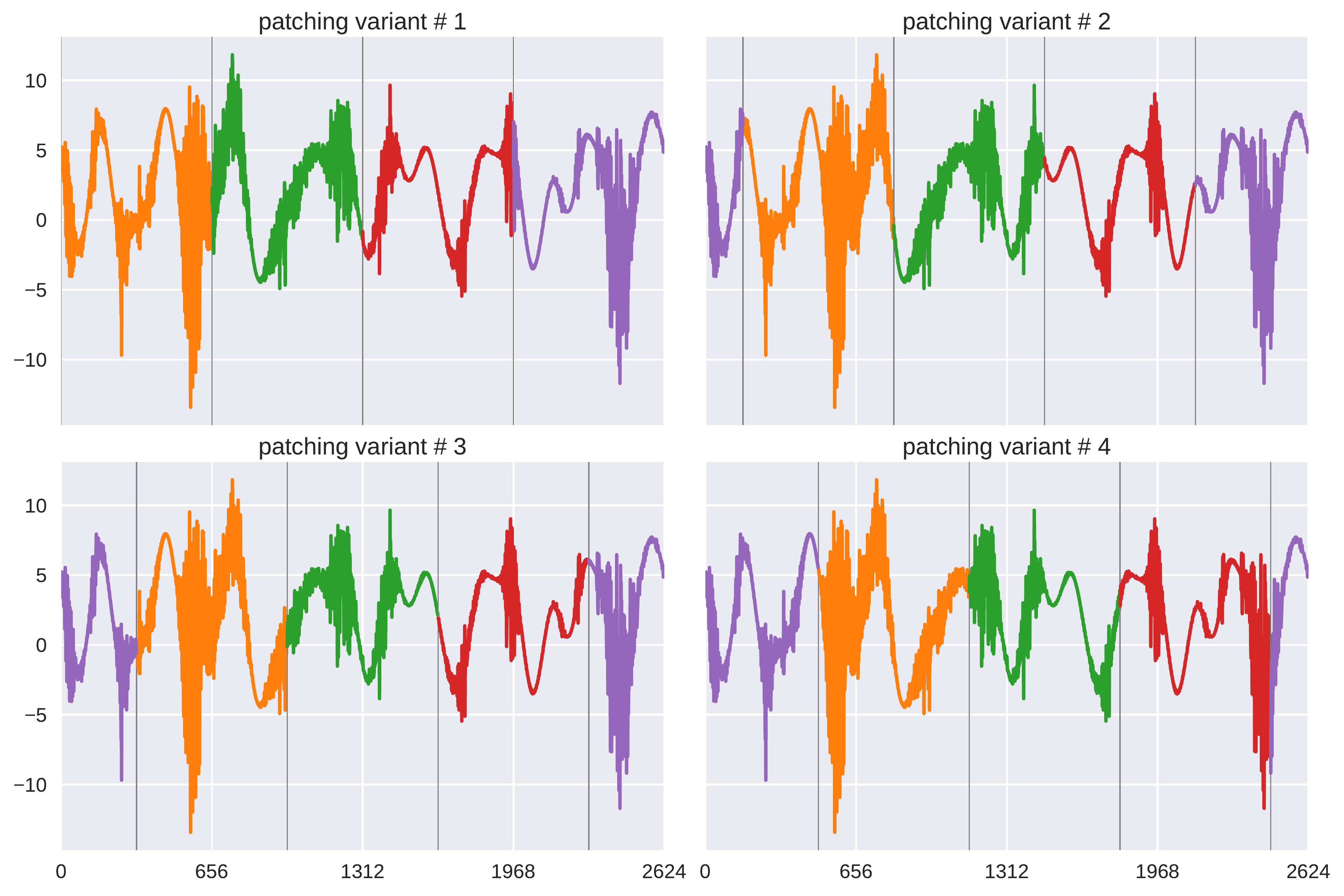}
    \caption{\label{fig:patches} Four patching variants of one instance of the 1D toy model in \cref{eqn:toymodel} with $J=64$ and $K=41$.}
\end{figure}

\begin{figure}[h]
    \centering
    \includegraphics[width=\textwidth]{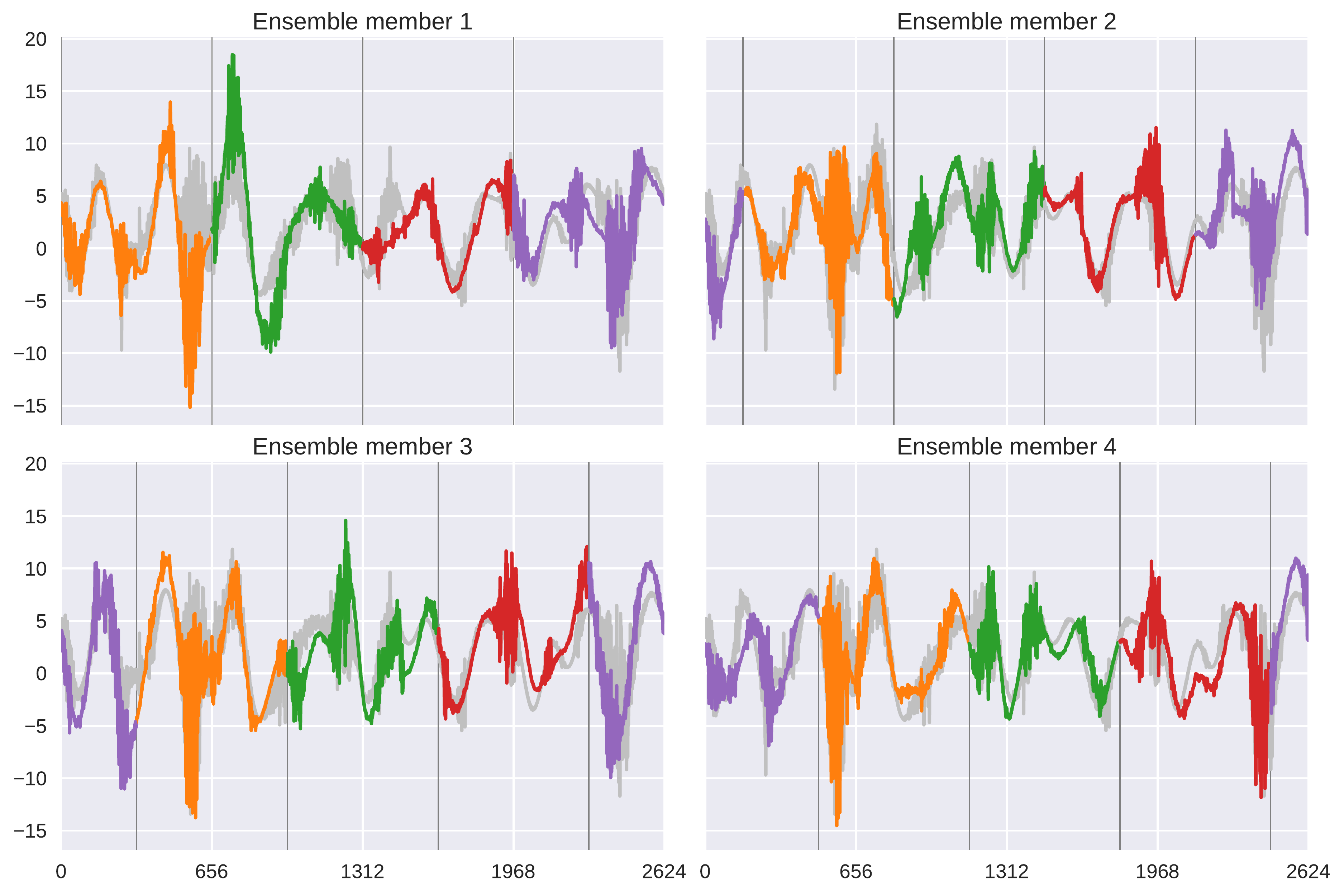}
    \caption{\label{fig:patches_ensemble} Example of four ensemble members generated by \cref{algo:p-gen}. The grey line is the forecast from which the patched analogs are derived.}
\end{figure}

As an example, consider implementing this to a state variable of the model \cref{eqn:toymodel} with $J=64$ and $K=41$, which has dimension $2624$.
\Cref{fig:patches} shows four distinct ways of splitting the state variable into four equal-sized patches of size $656$. 
Each color represents a single patch of size $1/4$ of the original domain size in each of the four patching variants, and the gray vertical lines mark the boundaries between two adjacent patches.
The patching variants use shifts of $0$, $164$, $328$, and $492$ units to the right. 
To retrieve the original variable, the patches are concatenated from orange to green to red to purple from left to right, then shifted by $0$, $164$, $328$, and $492$ units to the left.
Note that $656$ distinct patching variants exist for this particular example.
Lastly, \cref{fig:patches_ensemble} shows four distinct analog ensemble members generated from the same forecast.
Each ensemble member was created from the corresponding patch shift, as well as having had different perturbations of amplitude $r_z=0.75$ added while in their encoded states.
The different colors again represent the patches,the vertical lines represent the patch boundaries, and the grey waves in the background show the original forecast.
Next, we discuss how a general autoencoder can be used instead of a VAE within p-cAnEnOI.

\subsection{Constructing Analogs with General Autoencoders}\label{methods:patching_ae}
Recall that VAEs are autoencoders with extra conditions that force regularity in the latent space by adding an extra term in the loss function and require sampling via \cref{eqn:vae_sampling}.
In theory, the optimal encoder of a VAE should map from the space of the dataset at hand to the standard normal distribution, whereas the optimal encoder of a general autoencoder does not guarantee any special structure of the latent space.
This lack of regularity in the latent space makes general autoencoders a subpar candidate for generative modeling. 
For example, in \cref{algo:p-gen}, we add noise sampled from $N(\bb{0},r_z^2\bb{I})$ to the encoded state, since we can assume that the encoded state is a sample from $N(\bb{0},\bb{I})$.
For a general autoencoder there is no a priori way to predict the structure of the data distribution in latent space, and therefore no a priori way of knowing how much noise to add in latent space when contructing analogs.
One possible way to set the structure of the noise in latent space is to mimic the approach for a variational autoencoder.
Let $\bb{Z}$ be a random variable corresponding to the latent space distribution of a general autoencoder, and let the mean and covariance of $\bb{Z}$ be $\bb{\mu}$ and $\bb{\Sigma}$.
If $\bb{L}$ is the Cholesky factor of $\bb{\Sigma}$, then 
\begin{equation}
    \bb{Z}' = \bb{L}^{-1}(\bb{Z}-\bb{\mu})
\end{equation}
has mean $\bb{0}$ and covariance $\bb{I}$, analogous to the standard normal distribution in latent space that is targeted by a VAE.
Since we generate analogs using the VAE by adding noise sampled from $N(\bb{0},r_z^2\bb{I})$ to the encoded forecast state, it is natural for the general autoencoder to add noise sampled from $N(\bb{0},r_z^2\bb{I})$ to $\bb{Z}'$.
Let us call the encoded forecast state $\bb{z}\in\R^{\text{{\tt latentDim}}}$, which should be a sample from the distribution $\bb{Z}$.
Analogs in latent space $\bb{z}^{(i)}$ are constructed by transforming $\bb{z}$ to $\bb{z}'$, then adding noise, then converting back:
\begin{equation}
    \bb{z}^{(i)} = \bb{L}\left(\bb{L}^{-1}\left(\bb{z}-\bb{\mu}\right) + r_z\bb{\epsilon}^{(i)}\right) +\bb{\mu}= \bb{z} + r_z\bb{L\epsilon}^{(i)},\;\; \bb{\eps}^{(i)}\sim N(\bb{0},\bb{I}),\; i\in\{1,\cdots,N_e\}.
\end{equation}
Since the covariance $\bb{\Sigma}$ is not known a priori, it can be approximated by encoding many samples from the training set into the latent space and computing the mean vector $\hat{\mu}$, the covariance matrix, $\hat{\bb{\Sigma}}$, and the Cholesky factor such that $\hat{\bb{\Sigma}}\approx\hat{\bb{L}}\hat{\bb{L}}^{\top}$.
Finally, we can replace line 5 of \cref{algo:p-gen} with
\begin{equation}
    \bb{x}^{(j,i)} \gets d(\bb{z}_{\mu} + r_z\hat{\bb{L}}\eps^{(j,i)})\label{eqn:ae_cov}
\end{equation}
to add regular perturbations even when using a general autoencoder.
Outside the costs of training, this adds $N_e$ matrix-vector multiplications, each of size {\tt latentDim}, at every assimilation cycle as well as the initial costs of computing $\hat{\bb{\Sigma}}$, and its Cholesky factorization.
However, this will not significantly affect the computational cost since {\tt latentDim} is likely sufficiently small and these extra operations will not significantly inflate the leading order costs.
\subsection{Analysis: Impact of Patching on Covariance Matrix}\label{methods:cov}
There are two different effects at play in the structure of the background covariance matrix in p-cAnEnOI: sampling errors and the effect of patching.
To isolate the effect of patching we develop in this section a patched stochastic process model, and then derive an explicit formula for its covariance function.
Consider a stochastic process $Z(s)$ with finite second-order moments.
Without loss of generality we assume the mean is zero, and for simplicity of exposition we take $s\in\mathbb{R}$; extensions to higher spatial dimension are discussed at the end of the section.
Denote the covariance function of $Z$ by
\begin{equation}
    \text{Cov}[Z(s),Z(t)] = C(s,t).
\end{equation}
Next construct a `patched' stochastic process $Z_p(s)$ as follows.
Without loss of generality, let the patch width be 1.
A single sample of $Z_p$ is constructed by the following two-step procedure.
\begin{enumerate}
    \item Draw a shift $s_*$ from the uniform distribution on $[0,1)$ and define intervals $I_k = [s_*+k-1,s_*+k)$ for $k=-\infty,\ldots,\infty$.
    \item For each $k$ draw an independent sample of $Z(s)$ and then let $Z_p(s) = Z(s)$ for $s\in I_k$.
\end{enumerate}
In practice when constructing the ensemble members the patch boundaries are not drawn randomly; however, if a single ensemble member is drawn at random, the patch boundaries for that ensemble member can be considered as random variables.

To find the covariance function for the patched process $Z_p$ we will first find the covariance conditioned on the shift $s_*$, and then average over the uniform distribution on $s_*$.
If $s$ and $t$ are not in the same interval $I_k$ then the conditional covariance Cov$[Z_p(s),Z_p(t)]$ is zero.
If they are in the same interval, then the conditional covariance is the same as for $Z$, i.e. $C(s,t)$.
So the unconditional covariance function of $Z_p$ is $C(s,t)$ multiplied by the probability that $s$ and $t$ are both in the same interval, then averaged over the distribution of $s_*$.

Let $t=s+d$.
The probability that $s$ and $t$ are both in $I_{k}$ is zero if $|d|\ge1$.
If $|d|<1$, the probability that $s$ and $t$ are both in $I_{k}$ is 1 minus the probability that $s_*$ is in between them.
So
\[P[s,t\in I_{k}] = 1 - \min\{1,|s-t|\}.\]
The covariance function for the patched process is simply
\begin{equation}
    \text{Cov}[Z_p(s),Z_p(t)] = C(s,t)\left[1 - \min\left\{1,|s-t|\right\}\right].\label{eqn:PatchedCovariance}
\end{equation}
In higher dimensions $\bm{s}\in\mathbb{R}^d$ the form of the function changes slightly to reflect the general principle that the covariance function of the patched process is the covariance function of the original process multiplied by one minus the probability that there is at least one patch boundary between $\bm{s}$ and $\bm{t}$.

This is equivalent to using a distance-based localization (called `tapering' in the statistics literature) where the localization function is a `tent' function (the shape of the localization function changes in higher dimensions).
In order for the covariance function of the original and patched processes to remain similar, it is necessary to choose a patch size that is larger than the correlation length scale of the true process.
How much larger is, of course, situation dependent.
It is worth noting that the covariance matrix estimated from a patched ensemble can still exhibit spurious correlations at long ranges because of sampling errors, so localization methods still need to be used in patched cAnEnOI.
In a future work we plan to explore overlapping (or layering) the patches with smooth transitions across the finite overlap, rather than a complete independence across patch boundaries.

\section{Numerical Results}
\label{sec:NR}
We test our method on the 1D toy model described in \cref{methods:toymodel}, a multiscale modification introduced in \cite{grooms2015framework} to the Lorenz '96 model of \cite{lorenz1996predictability}, and recently used as a data assimilation test model in \cite{grooms2020analog,robinson2020hybrid}.
First, we generated a library of snapshots by numerically solving \cref{eqn:toymodel} via {\tt ode45} in Matlab until final time $t=200$, each snapshot covered the full spatial domain.
Since $0.2$ time units is about one day of atmospheric dynamics according to \cite{lorenz1996predictability}, the simulation is run almost to $3$ years.
This library contains $270,576$ samples, $25\%$ of which were reserved for validation. 
After separating this library into a training set and a validation set, every element in the library was partitioned into smaller patches as our goal was to test the patch version of cAnEnOI.
The patch sizes ranged from $1/32$ to $1/4$ of the original domain, and thus we created in total five libraries of the same data but with different data sizes.
We elaborate on the details of training the VAEs and using them in p-cAnEnOI in \cref{NR:vae} and show the results of using general autoencoders in \cref{NR:ae}.

\subsection{p-cAnEnOI with VAE}\label{NR:vae}
A main component of our method outlined in \cref{methods:patching} is the use of a machine learning model as a generative tool to create artificial ensemble members.
In this section, we share the architectures of the specific machine learning models we used, how these models were trained, and the performance of these models with respect to standard machine learning evaluation metrics.
Ultimately, we are most interested in how the artificial ensemble members impact the performance of the data assimilation task and we discuss the relationship between the quality of the machine learning models and the accuracy of the data assimilation in \cref{NR:DA}.
We investigated autoencoders and variational autoencoders (see \cite{Kingma2019}) and trained several models using the library of patches of snapshots of the 1D toy model discussed in \cref{methods:toymodel}.\par


\subsubsection{VAE Training}\label{NR:train_vae}
We used the variational autoencoder structure used in \cite{grooms2020analog}, \cref{model:e_correct},\cref{model:d_correct}, for patches with sizes $1/4$, $1/8$, and $1/16$ of the original variable.
\Cref{model:e_correct} reduces the dimension of the data to $1/4$ of the input size with two max pooling layers (each pooling layer halves the dimension), and the fully connected layer reduces it further to two outputs that are each approximately $1/6$ of the input size. 
These two outputs $z_{\mu}$ and $z_{\log(\bb{\Sigma})}$ are then combined via \cref{eqn:vae_sampling}, completing the encoding into the latent space. 
Finally, the original dimensionality is recovered by \cref{model:d_correct}. 
We used \cref{model:e_correct_32} for the $1/32$ patches because each sample was not divisible by $4$ and therefore could not undergo the quarter reduction of the dimensions with the two max pooling layers. 
Therefore, \cref{model:e_correct_32} only has one max pooling layer and uses the fully connected layer to reduce the dimension to approximately $1/6$ of the input size.
Similarly, \cref{model:d_correct_32} only has one transposed convolutional layer with stride $(2,1)$ as opposed to the two layers in \cref{model:d_correct}.\par 

The VAEs were trained using the loss function as defined in \cref{methods:ML}. 
This loss term includes the reconstruction error and the KL-divergence of the training set.
Due to risk of overfitting to the training data, a validation set is used to compute an unbiased measure of the quality of a model while being trained.
We examine only the reconstruction error portion of the loss on the validation set.
Suppose that there are $N$ elements of a dataset where each element is of {\tt dsize} dimensions. 
Then, the reconstruction error (RE) is defined via \cref{eqn:smm},
\begin{equation}
    \text{RE} = \left(\frac{1}{N}\sum_{i=1}^N\frac{1}{{\tt dsize}}\sum_{j=1}^{{\tt dsize}}(\bb{x}_i^{(j)}-d(e(\bb{x}_i^{(j)})))^2\right)^{1/2},
    \label{eqn:smm}
\end{equation}
where $e:\R^{{\tt dsize}}\rightarrow\R^{{\tt latentDim}}$ and $d:\R^{{\tt latentDim}}\rightarrow\R^{{\tt dsize}}$ are some encoder and decoder, and $\bb{x}_i^{(j)}$ is the $j^{th}$ element of the $i^{th}$ sample.
Recall that we used one library of snapshots to generate four more libraries of the same snapshots partitioned into the four different patch sizes.
We evaluate the quality of our VAEs via \cref{eqn:smm}, which averages over a dataset with $N$ elements of dimension ${\tt dsize}$ and applies the square root at the very end to ensure uniformity across the four different libraries.
Note that $N\times {\tt dsize}$ for the validation set for all five libraries remains at the same value, $44,374,464$.\par
\begin{figure}[t]
	\centering
	\includegraphics[width=\textwidth]{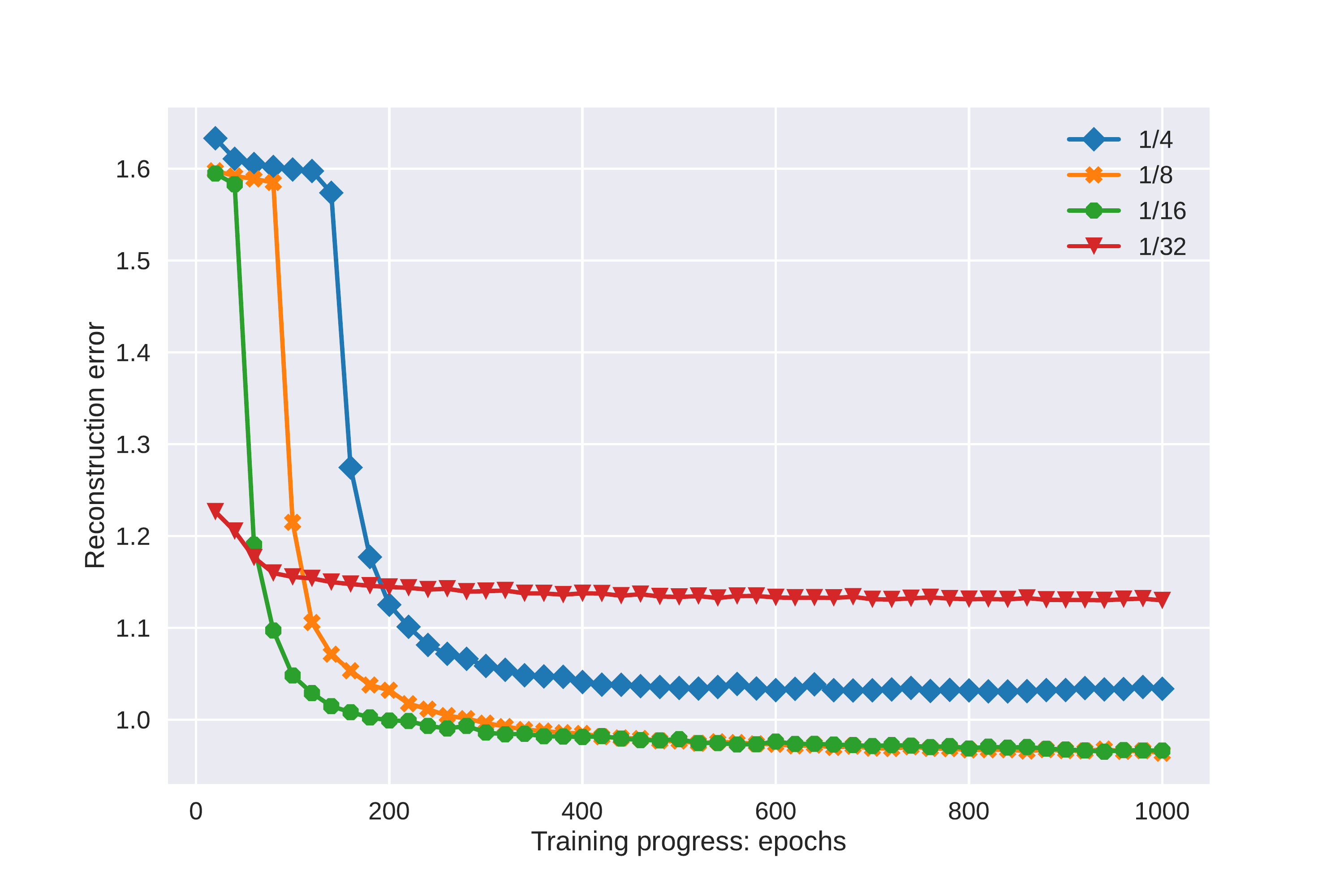}
	\caption{\label{fig:re_epochs} Averaged reconstruction error computed via \cref{eqn:smm} on the validation set shown every 20 epochs during training. }
\end{figure} 

In \cref{fig:re_epochs}, we show the averaged RE for the validation set every $20$ epochs until the $1,000^{th}$ for the patch sizes $1/4$, $1/8$, $1/16$, and $1/32$ while training the VAE architectures [\cref{model:e_correct}, \cref{model:d_correct}] and [\cref{model:e_correct_32},\cref{model:d_correct_32}].
We can see that the VAEs for all patch sizes experience a significant drop in the reconstruction error in the first $200$ epochs.
The drop happens sooner at smaller patch sizes, to the extent that at the smallest patch size it happens before the first point on the plot, at 20 epochs.
If we exclude the smallest patch size, then we observe that the smaller patch sizes are able to achieve smaller reconstruction errors.  
Furthermore, the VAE reconstruction errors for the $1/4$ and $1/32$ patch sizes seem to have reached their minima within the first $500$ epochs, whereas the errors for the other two patch sizes could potentially decrease further.\par
\begin{figure}[b!]
	\centering
	\includegraphics[width=\textwidth]{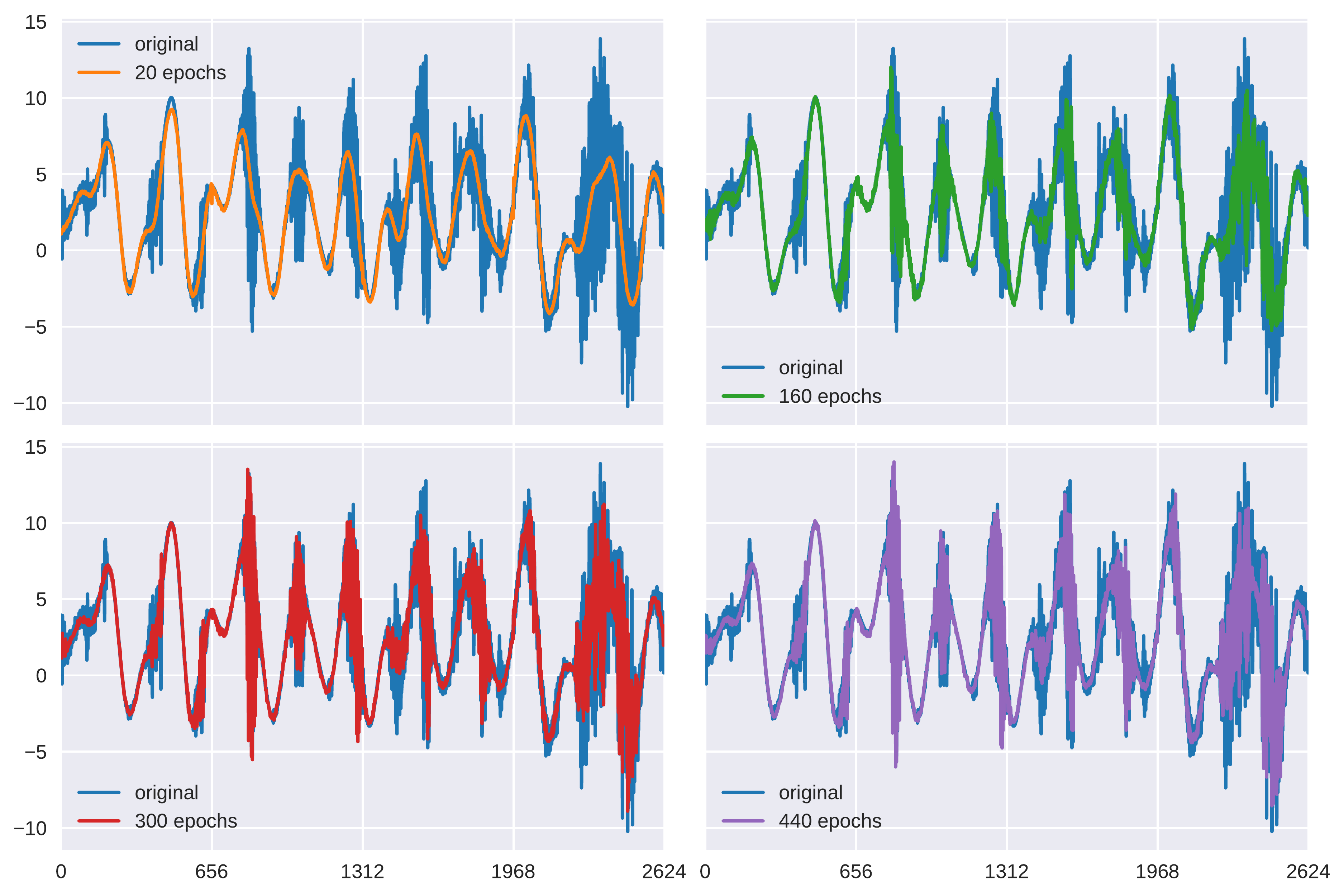}
	\caption{\label{fig:vae_training} A sample in the validation set is shown with its reconstruction after $20$, $160$, $300$, and $440$ epochs of training. The patch size is 1/4 of the domain size.}
\end{figure}

Next, \cref{fig:vae_training} shows how the quality of the $1/4$ patch VAE improves with more training for a single sample. 
The original sample is shown in blue, and the other colors show the reconstruction of that sample with the VAE after $20$, $160$, $300$, and $440$ epochs of training.
Corresponding to the large drop in the reconstruction error in the $1/4$ patch between the $140^{th}$ and the $180^{th}$ epoch in \cref{fig:re_epochs}, the top right panel shows the most visibly recognizable improvement at epoch $160$, while the bottom two plots certainly show continued progress.
Combining \cref{fig:re_epochs} and \cref{fig:vae_training} we conclude that the VAE rapidly learns to reproduce the large scales, and then after sufficient training it begins to learn the small scales; the rapid drop in reconstruction error evident in \cref{fig:re_epochs} is evidently associated with learning the small scales.

\begin{figure}[h]
	\centering
	\includegraphics[width=.49\textwidth]{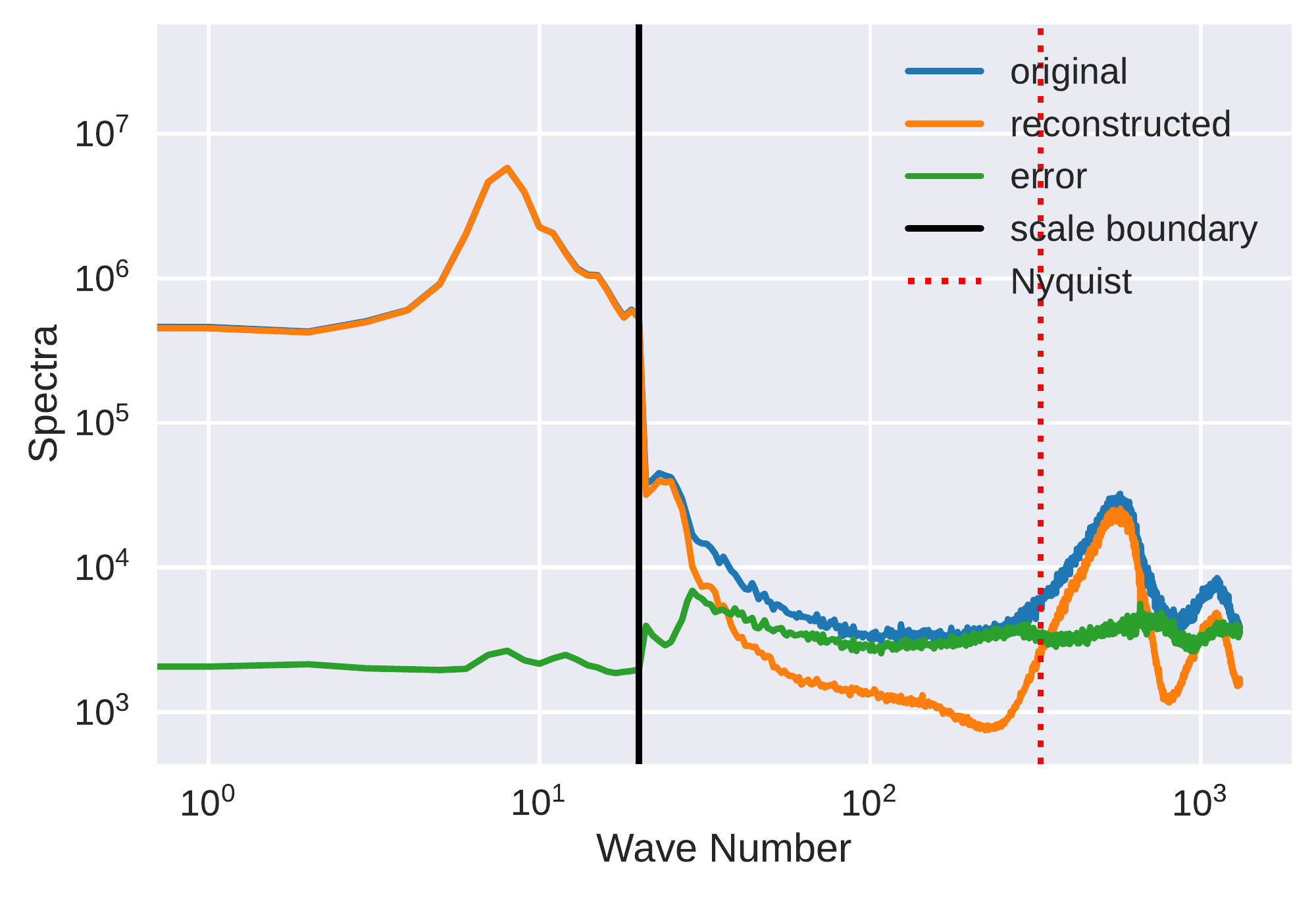}
	\includegraphics[width=.49\textwidth]{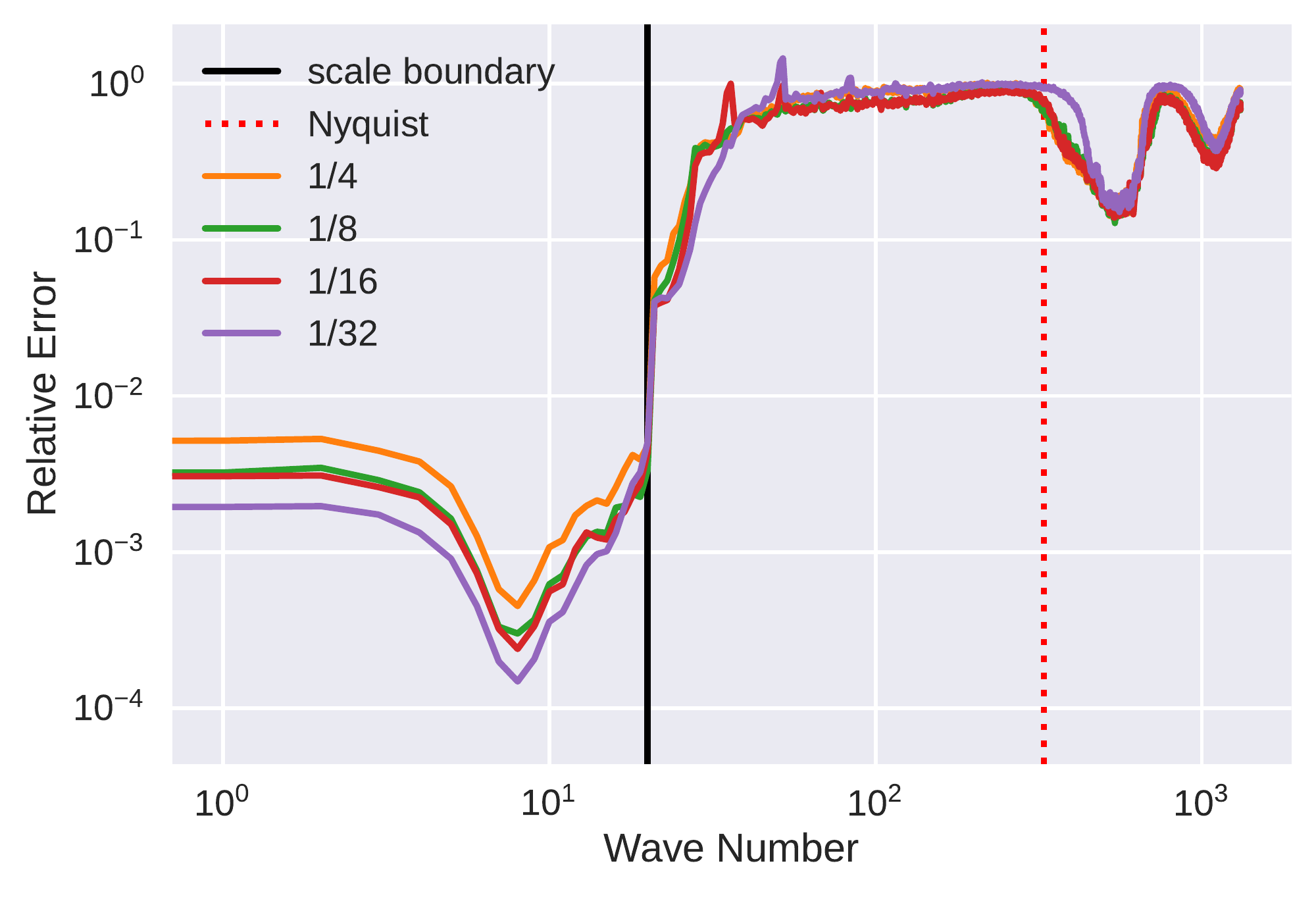}
	\caption{\label{fig:spectra} Left: The Fourier spectra of a sample from the validation set, its reconstruction using a full domain VAE, and the absolute error. Right: The relative error in the Fourier spectra of the same sample VAEs with various patch sizes. The solid black line indicates the cutoff scale between the 41 large-scale modes and the remaining small-scale modes, while the dashed black line indicates the Nyquist wavenumber of the observing system.}
\end{figure}

Finally, \cref{fig:spectra} shows the relative errors of the spectra for VAEs for the four patch sizes for $500$ epochs on a sample from the validation set.
There is a sharp increase in the error exactly at the scale boundary, which suggests that the large scale dynamics are well emulated by the all four VAEs. The smaller patch VAEs actually show lower errors in the large scales, but the errors are indistinguishable for the large wave numbers. 
In conclusion, the VAEs for the larger patch sizes incur more loss in reconstruction in comparison to the VAEs for smaller patch sizes when trained for the same number of epochs.

\subsubsection{Data Assimilation Results}\label{NR:DA}
The reference simulation is initialized with standard normal noise and run until $t=9$, by which the statistical equilibrium has been reached.
The data assimilation starts from the state at $t=9$, and the reference is simulated for 73 additional time units with time-step size 0.2, which corresponds to one `day' according to \cite{lorenz1996predictability}. 
The observations are taken at every 4th spatial node at $0.2$ time units, and the observation errors are sampled from $N(0,1/2)$.
There are a total of $73\div0.2=365$ assimilation cycles, of which the first 73 are discarded as the burn-in period of the data assimilation. 
The root mean squared error (RMSE, see \cref{eqn:DA_RMSE}) between the reference simulation and the analysis mean for the latter 292 assimilation cycles quantify the performance of p-cAnEnOI for the specific set of parameters used. 
Each combination of the three parameters undergoes 8 trials to account for statistical variability and we summarize the performance with the mean and standard error across these 8 trials.
Note that several experiments from \cite{grooms2020analog} were continued to 1,000 assimilation cycles with no change in the performance, indicating that 8 independent trials with 365 cycles each is sufficient to assess the performance of the different methods. \par

These settings are set identical to the experiments in \cite{grooms2020analog} so the patched constructed analog method can be compared to those.
The data assimilation performance was measured using RMSE \cref{eqn:DA_RMSE}, which measures the error between reference simulation and the ensemble mean after assimilating observations.
It was found in \cite{grooms2020analog} that a 200 sized ensemble showed little improvement over an ensemble size of 100, and therefore $N_e=100$ was used for all methods, ESRF, EnOI, AnEnOI, and cAnEnOI. 
EnOI provides a lower bound as expected, and AnEnOI and cAnEnOI result in lower RMSE than EnOI. 
The mean RMSE for cAnEnOI is actually lower than the mean RMSE from ESRF but the 8 independent trials were used to show that this difference is not statistically significant.
We summarize the relevant results from \cite{grooms2020analog} in \cref{table:summaryGrooms}. 
\begin{table}[h]
\centering
\begin{tabular}{ |c|c|c|c|c|c| }
     \hline
      & ESRF & EnOI & AnEnOI & cAnEnOI \\ \hline 
      DA RMSE & $1.35$ & $2.27$ & $2.01$ & $1.30$\\
     \hline
\end{tabular}
\caption{\label{table:summaryGrooms} Summary of various analog data assimilation methods from \cite{grooms2020analog} with optimal data assimilation parameters and $r_z$ for cAnEnOI.}
\end{table}

Our new method, p-cAnEnOI, uses a generative machine learning model to construct patched-analog ensemble members for use within EnOI. 
The relevant tunable parameters include: $r_z$, the amplitude of noise added in the latent space when generating articial ensemble members via \cref{eqn:gen_en}; $f$, the forecast spread of the generated ensemble; $L$, the localization radius to zero out spuriously high correlations across far distances. 
We searched through the parameter spaces of $r_z$, $f$, and $L$ for p-cAnEnOI with VAEs for each of the patch sizes $1/4$, $1/8$, $1/16$, and $1/32$, with 8 trials for each combination.
The VAEs used the architecture [\cref{model:e_correct}, \cref{model:d_incorrect}] for patch sizes $1/4$, $1/8$, $1/16$ and [\cref{model:e_correct_32}, \cref{model:d_incorrect_32}] for patch size $1/32$. \par

\begin{figure}[h]
	\centering
	\includegraphics[width=\textwidth]{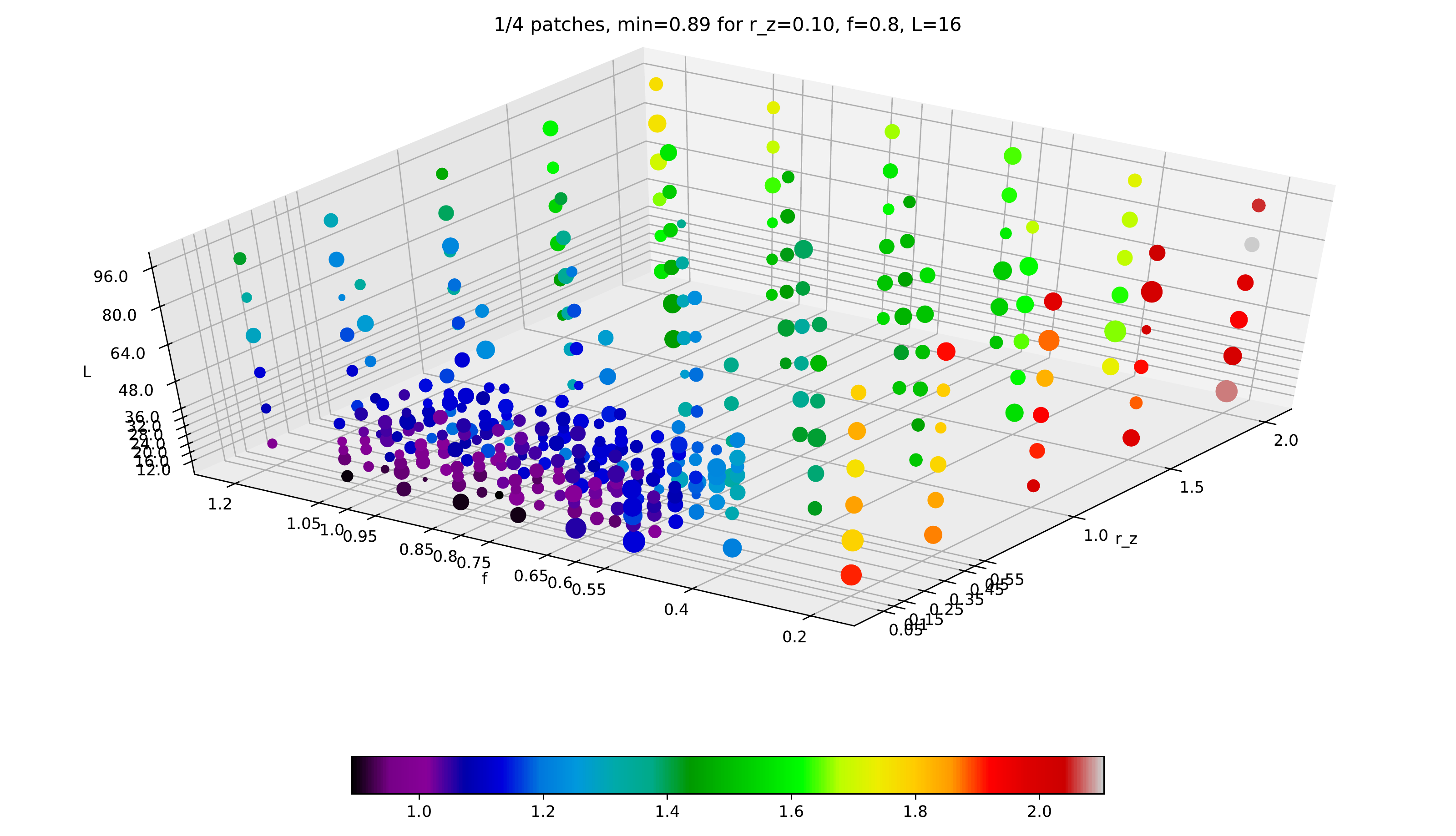}
	\caption{\label{fig:psweep} The color and the size of the points denote the computed mean and the standard error of the RMSE across 8 trials and their locations mark the combination of the three relevant parameters: $r_z$, the amplitude of noise added in latent space to generate artificial ensemble members; $f$, the forecast spread; and $L$, the localization radius. Larger points imply larger RMSE, and smaller points, smaller RMSE. This plot shows the data for the $1/4$ patch size p-cAnEnOI. 
	}
\end{figure}

\Cref{fig:psweep} shows the results of the parameter sweep for the p-cAnEnOI that uses patch size $1/4$.
The color of each point represents the average RMSE value over 8 trials for a specific combination of the three parameters $r_z$, $f$, and $L$, and we see that the parameters significantly influence the results. 
When $r_z=0.10$, $f=0.8$, and $L=16$, the optimal RMSE of $0.89$ is achieved for the p-cAnEnOI scheme with $100$ patched-analog ensemble members, which is significantly lower than that achieved by the global cAnEnOI and the ESRF. 
This improvement could suggest that the localization associated with patching (\cref{methods:cov}) is more effective than the traditional localization described by \cref{eqn:loc}. \par 
 
\Cref{table:summary} shows the optimal set of parameters for each of the patch sizes used, the corresponding mean and standard errors of the RMSE, and the mean reconstruction error of the validation set.
If we exclude the column with the global cAnEnOI, there is a clear trend that shows larger patch sizes achieving the best optimal RMSE.
While the optimal $f$ value stays within $[0.70,0.90]$ for all variants, the localization radius $L$ and amplitude of noise in latent space, $r_z$, vary quite a bit.
The global cAnEnOI from \cite{grooms2020analog} bucks this trend, but it might be that the VAE used in the global patch has not yet been trained to its optimal performance level, because it is significantly more expensive to train than the VAEs used for the smaller patches.

The variation in optimal localization radius may be explained by reference to the results in \cref{methods:cov}.
Patched analogs have an effect similar to localizing the covariances.
At large patch sizes the amount of localization needed is comparable to an un-patched method; at small patch sizes the effect of patching already significantly reduces short-range correlations, but some localization is still needed to remove spurious correlations at long ranges.

The larger patch sizes ($1/4$ and $1/8$) benefit from smaller $r_z$, whereas the smaller two patch sizes ($1/16$ and $1/32$) prefer larger $r_z$'s.
This sensitivity to $r_z$ might be explained as follows.
Recall that VAEs are designed so that the encoder is a mapping from the space of instances of input variables to the latent space, which is forced towards the multivariate standard normal distribution, $N(\bb{0}, \bb{I})$. 
It is possible that the VAEs for the two larger patch sizes encode into latent spaces having enough dimensions to allow for preserving deeper complexities than the VAEs for the smaller patch sizes.
If so, even a small amplitude noise added in those more complex latent spaces could generate artificial ensemble members that exhibit enough diversity, whereas a larger amplitude is necessary to generate distinct enough ensemble members for the VAEs with smaller latent space dimensions.\par

\begin{table}[h]
\centering
\begin{tabularx}{\textwidth}{|Y|c|c|c|c|c|}
     \hline
     Patch size & 1 & 1/4 & 1/8 & 1/16 & 1/32\\
     \hline
     $L$ & 40 & 16 & 12 & 24 & 36\\
     $f$ & 0.70 & 0.80 & 0.70 & 0.80 & 0.90\\
     $r_z$ & 0.70 & 0.10 & 0.05 & 0.25 & 0.45\\ \hline
     \makecell[c]{DA RMSE:\\ \cref{model:d_incorrect}, \ref{model:d_incorrect_32}}& n/a & $0.89\pm0.01$ & $0.95\pm0.01$ & $1.13\pm0.02$ & $1.18\pm 0.03$\\
    \hline
    \makecell[c]{DA RMSE:\\ \cref{model:d_correct}, \ref{model:d_correct_32}}
   & $1.30\pm0.05$& $0.88\pm0.01$ & $0.93\pm0.02$ & $1.17\pm0.02$ & $1.39\pm 0.03$\\
    \hline
    RE: \cref{model:d_incorrect}, \ref{model:d_incorrect_32}& n/a & $1.03$ & $0.97$ & $0.96$ & $0.78$\\
    \hline
   RE: \cref{model:d_correct}, \ref{model:d_correct_32}& $1.26$& $1.03$ & $0.99$ & $0.97$ & $1.13$\\
    \hline
\end{tabularx}
\caption{\label{table:summary} Summary of optimal parameters for p-cAnEnOI for patch sizes $1/4$, $1/8$, $1/16$, and $1/32$, as well as the global cAnEnOI, and the corresponding mean and standard error of the data assimilation RMSE and reconstruction error (RE).}
\end{table}

\begin{figure}[h]
	\centering
	\includegraphics[width=\textwidth]{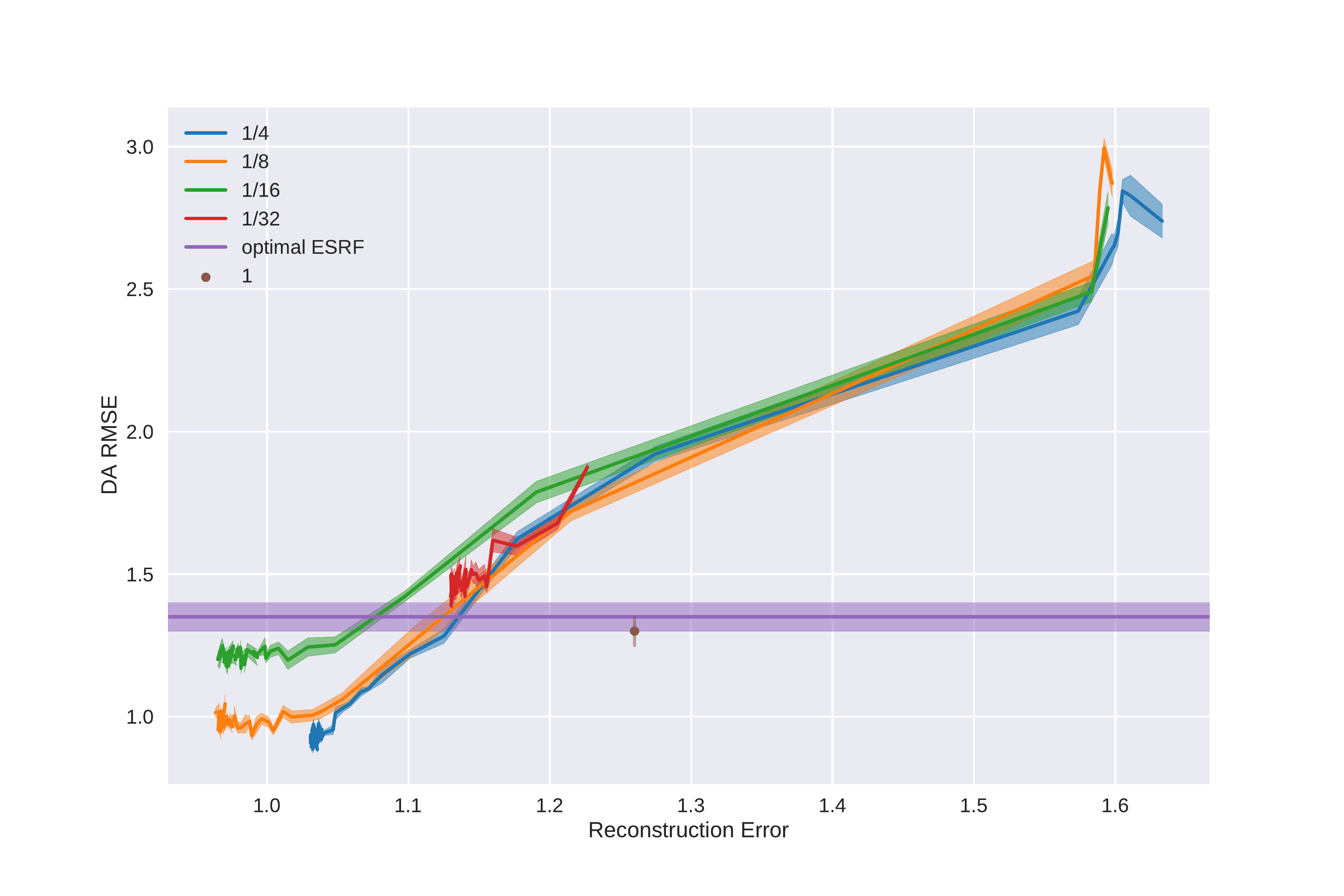}
	\caption{\label{fig:re_da} Averaged reconstruction error of VAEs in training over validation data set compared against the resulting mean and standard error of DA RMSE. The shaded regions show the range of the standard error around the mean values.}
\end{figure}

To explore the relationship between the accuracy of the VAEs as measured by the reconstruction error \cref{eqn:smm} and the performance of the corresponding p-cAnEnOI, we return to the set of VAEs discussed in \cref{NR:train_vae} which used [\cref{model:e_correct}, \cref{model:d_correct}] for patch sizes $1/4$, $1/8$, $1/16$ and [\cref{model:e_correct_32}, \cref{model:d_correct_32}] for patch size $1/32$.
The VAEs were saved every 20 epochs during training to track the progress of the quality of the VAEs with respect to the reconstruction error (see \cref{eqn:smm}) over the validation set.
Using the set of optimal parameters $(L,f,r_z)$ found using \cref{model:d_incorrect,model:d_incorrect_32} (and shown in \cref{table:summary}) for each patch size, we computed the data assimilation mean RMSE over 8 trials for each of the VAEs saved throughout the first 500 epochs of training.
\Cref{fig:re_da} shows the trend between the reconstruction error \cref{eqn:smm} and the mean RMSE \cref{eqn:DA_RMSE} for each patch size, and the optimal RMSE out of the 50 saved VAEs for each patch size are reported in \cref{table:summary}.\par

As the reconstruction error decreases with more training, the quality of the data assimilation also improves as is shown by lower mean RMSE values.
In fact, there is little distinction between p-cAnEnOI with patch size $1/4$, $1/8$, and $1/16$ for VAEs that yielded reconstruction error between $1.2$ to $1.5$, other than that the $1/8$ patch size has a slightly steeper slope.
The $1/16$ variant splits off from the other three and a higher mean RMSE is computed where the reconstruction error ranges approximately from $0.95$ to $1.2$. 
The $1/8$ variant produces mean RMSE close to 1 even when the reconstruction error continues to decrease from about 1.3 to 0.95, and the $1/4$ variant produces mean RMSE that dips slightly below 1 despite the reconstruction error stagnating around 1.3.
The reconstruction error for $1/32$ cannot improve after $~1.12$ and the RMSE similarly is restricted to range from $1.35$ to $1.5$, which indicates that the small latent space of this VAE cannot represent the full complexity of the dynamics of this model.
Overall, there is a clear linear relationship between the reconstruction error and the mean RMSE of the data assimilation performance.
The slight disparities between these p-cAnEnOI variants may be attributed to the varying patch sizes and the second loss term, the KL-divergence. 
For example, it is possible that the $1/16$ p-cAnEnOI yields a higher RMSE than p-cAnEnOI with patch sizes $1/4$ and $1/8$ despite having the same reconstruction error of 1.03 because that latent space of the $1/16$ patch vae is less regular in comparison to the other two.
Or, it could be that that $1/16$ patch VAE fails to capture the large scale motions accurately enough since the smallest large scale wave has length approximately $1/8$ of the domain.\par

In \cref{table:summary}, we see that the two sets of VAEs (with different decoders) for each patch size $1/4$, $1/8$, and $1/16$ produce RMSE and RE that are very similar, but the two VAEs for patch size $1/32$ are quite different from each other.
The mean RE for the VAE with \cref{model:d_incorrect_32} for patch size $1/32$ yields the smallest reconstruction error out of all 9 VAEs, but still yields the second worst RMSE.
This strongly suggests that too small of a patch size loses too much information and cannot act as a good generative model for constructing analogs.

Most importantly, we observe that the p-cAnEnOI schemes with patch sizes (strictly) larger than $1/32$ and the global cAnEnOI outperform the optimal ESRF although ESRF could probably do better with more sophisticated localization and inflation.
While we did not measure the exact computation cost for these variants, implementation of ESRF is far more expensive in the forecasting step, which often accounts for the leading order computational expense.

\subsection{p-cAnEnOI with a general autoencoder}\label{NR:ae}
A main component of VAEs is the effort to regularize the latent space by encouraging it to look like the standard normal distribution, and is done by \cref{eqn:vae_sampling} and the inclusion of the KL-divergence in the loss term as is discussed in \cref{methods:VAE}. 
To provide a basic comparison, we used \cref{model:e_correct} that only outputs one variable of size {\tt latentDim} at the last layer, and \cref{model:d_incorrect} exactly as is to form a simple autoencoder for patch size $1/8$ and trained it only with the reconstruction error as the loss term.
After a week of training, the reconstruction error computed via \cref{eqn:smm} using the autoencoder on the validation set yielded 0.92, a significantly lower value than the two corresponding VAEs (a week of training with \cref{model:d_incorrect} and 1000 epochs with \cref{model:d_correct}).
The computed RMSEs had mean $0.96$ and standard error $0.02$, which shows that a general encoder can produce similar results as VAE within p-cAnEnOI with the additional step described in \cref{methods:patching_ae}.

\section{Conclusion}\label{sec:conclusion}
Using analogs and constructed analogs in EnOI (see \cite{grooms2020analog} for details) is a recent development in the growing field of using machine learning techniques for data assimilation.
AnEnOI and cAnEnOI alleviate the computational challenge of having to forecast large ensembles in EnKF and its variants, but improves upon EnOI by inserting time dependency to the ensemble such that the background covariance depends on the current forecast.
A limitation of cAnEnOI from \cite{grooms2020analog} is the need to train a generative machine learning model to construct samples of the entire model state, which can include billions of variables for geophysical applications.
We have introduced p-cAnEnOI which uses VAEs and general autoencoders to generate patches of analogs that form whole analogs when assembled.
This new scheme addresses the scalability of implementing cAnEnOI in geophysical applications in several ways: 1) Dividing the spatial domain into smaller subsets reduces the complexity of the machine learning model to be trained and encourages robust training with fewer parameters learned with more training samples available; 2) Multiple analog patches can be simultaneously generated in a parallel process then combined together, allowing for speed-ups. \par

Our numerical experiments were designed to test if and how the patch size affects the data assimilation performance, as well as how the accuracy of the generative model affects the data assimilation performance.
First, we trained VAEs that learned patches of instances of the toy model (defined in \cref{eqn:toymodel}) with sizes ranging from $1/4$, $1/8$, $1/16$, and $1/32$ of the spatial domain.
Then, we solved for the optimal set of three tunable parameters of the data assimilation system ($f$, forecast spread; $r_z$, the amplitude of noise added in latent space; and $L$, the localization radius) by running 8 trials of p-cAnEnOI for ranges of those parameters for each of the four patch sizes. 
We used the root mean squared error between the reference simulation and the analysis mean as defined in \cref{eqn:DA_RMSE} to measure the quality of data assimilation, and found that using larger patches led to better results. 
While the optimal forecast spread was similar across the different patch sizes, the localization radius and $r_z$ varied quite a bit.
Notably, the larger patches required smaller localization radii unlike smaller patches that naturally perform some localization. 
However, patch sizes that are large enough to capture the dominant large-scale structures in the data have similar optimal localization radii and small $r_z$, and variations in these parameters appear only for patch sizes that are sub-optimal. \par

All of the p-cAnEnOI variants produced better data assimilation RMSE than cAnEnOI, AnEnOI, EnOI, and ESRF.
Given that the large patches produce better results, it may be surprising that p-cAnEnOI performed better than cAnEnOI. 
Since we found that the normalized reconstruction errors (see \cref{eqn:smm}) of VAEs for the small patch sizes were lower, we sought to find a relationship between the reconstruction error and the data assimilation RMSE.
In this second experiment, we saved the VAEs at various stages during training and tested them within p-cAnEnOI with the optimal parameters found from the previous experiment.
As expected, more training produced better reconstruction errors, which then led to better data assimilation results. 
However, when comparing p-cAnEnOI with similar reconstruction errors, we found that the variants that use larger patches still performed marginally better. 
The lower than expected performance of global cAnEnOI from \cite{grooms2020analog} might thus be due to its use of an incompletely trained VAE.
Training the global VAE to its optimal performance is significantly more expensive than training a VAE for local patches. \par 

Lastly, we tested the use of general (non-variational) autoencoders in p-cAnEnOI. 
We implemented the method outlined in \cref{methods:patching_ae} for the $1/8$ patch with the parameters found for that patch size in the first experiment. 
The data assimilation RMSE averaged over 8 trials for this general autoencoder version produced results not statistically different from the VAE version.
This is a promising result, since it suggests that autoencoders with simpler architecture than VAEs, and which are therefore easier to train, might perform as well as VAEs for this task.\par 
In conclusion, constructing patched analogs shows encouraging results that prompt future studies of applying cAnEnOI and p-cAnEnOI to more realistic models. 
Both the patching scheme and the use of general autoencoders reduce the cost of training the autoencoder compared to global cAnEnOI.
In future work, we will address what patching scheme to use in non-periodic domains with boundaries. 
We will also investigate overlapping the patches with a smooth transition region rather than a sharp patch boundary in order to mitigate the across-patch discontinuities.
Since convolutional neural networks can be used for categorization, we believe they may be able to learn diverse sets of samples that can encompass the dynamics near the boundaries as well as away from the boundaries. 
As there is an abundance of historical datasets of geophysical systems, our study of utilizing these in an efficient way to improve data assimilation methods is meaningful.

\bibliography{library,sans_library,newer}
\appendix
\section{Model Architectures}
\begin{encoder}[Used for global and $1/4$, $1/8$, $1/16$ patches]\label{model:e_correct}
\begin{enumerate}
    \item A convolutional layer with 3 filters of size (3,1) and an $\mathrm{elu}$ activation.
    \item A convolutional layer with 9 filters of size (3,1) and an $\mathrm{elu}$ activation.
    \item A convolutional layer with 27 filters of size (3,1) and an $\mathrm{elu}$ activation.
 	\item A max pooling layer with pool size (2,1).
	\item A convolutional layer with 27 filters of size (3,1) and an $\mathrm{elu}$ activation.
	\item A convolutional layer with 27 filters of size (3,1) and an $\mathrm{elu}$ activation.
	\item A max pooling layer with pool size (2,1).
	\item A convolutional layer with 27 filters of size (3,1) and an $\mathrm{elu}$ activation.
	\item A convolutional layer with 27 filters of size (3,1) and an $\mathrm{elu}$ activation.
	\item A max pooling layer with pool size (2,1).
	\item A fully connected layer with two outputs $z_{\mu}$ and $z_{\log(\bb{\Sigma})}$, each of size {\tt latentDim}.
\end{enumerate}
\end{encoder}
\begin{encoder}[Used for $1/32$ patches]\label{model:e_correct_32}
\begin{enumerate}
    \item A convolutional layer with 3 filters of size (3,1) and an $\mathrm{elu}$ activation.
    \item A convolutional layer with 9 filters of size (3,1) and an $\mathrm{elu}$ activation.
    \item A convolutional layer with 27 filters of size (3,1) and an $\mathrm{elu}$ activation.
 	\item A max pooling layer with pool size (2,1).
	\item A convolutional layer with 27 filters of size (3,1) and an $\mathrm{elu}$ activation.
	\item A convolutional layer with 27 filters of size (3,1) and an $\mathrm{elu}$ activation.
	\item A max pooling layer with pool size (2,1).
	\item A fully connected layer with two outputs $z_{\mu}$ and $z_{\log(\bb{\Sigma})}$, each of size {\tt latentDim}.
\end{enumerate}
\end{encoder}
\begin{decoder}[Used for global and $1/4$, $1/8$, $1/16$ patches]\label{model:d_correct}
\begin{enumerate}
    \item A fully connected layer, which is then reshaped to 27 channels, followed by an $\mathrm{elu}$ activation.
 	\item A transposed convolutional layer with 27 filters of size (3,1), stride of (2,1), and an $\mathrm{elu}$ activation.
 	\item A convolutional layer with 27 filters of size (3,1), unit stride, and an $\mathrm{elu}$ activation.
 	\item A transposed convolutional layer with 9 filters of size (3,1), stride of (2,1), and an $\mathrm{elu}$ activation.
 	\item A convolutional layer with 9 filters of size (3,1), unit stride, and an $\mathrm{elu}$ activation.
 	\item A convolutional layer with 1 filter of size (3,1), unit stride, and an $\mathrm{elu}$ activation.
\end{enumerate}
\end{decoder}

\begin{decoder}[Used for $1/32$ patches]\label{model:d_correct_32}
\begin{enumerate}
    \item A fully connected layer, which is then reshaped to 27 channels, followed by an $\mathrm{elu}$ activation.
 	\item A transposed convolutional layer with 9 filters of size (3,1), stride of (2,1), and an $\mathrm{elu}$ activation.
 	\item A convolutional layer with 9 filters of size (3,1), unit stride, and an $\mathrm{elu}$ activation.
 	\item A convolutional layer with 1 filter of size (3,1), unit stride, and an $\mathrm{elu}$ activation.
\end{enumerate}
\end{decoder}

\begin{decoder}[Used for global and $1/4$, $1/8$, $1/16$ patches]\label{model:d_incorrect}
\begin{enumerate}
    \item A fully connected layer, which is then reshaped to 27 channels, followed by an $\mathrm{elu}$ activation.
 	\item A transposed convolutional layer with 27 filters of size (3,1), stride of (2,1), and an $\mathrm{elu}$ activation.
 	\item An $\mathrm{elu}$ activation.
 	\item A transposed convolutional layer with 9 filters of size (3,1), stride of (2,1), and an $\mathrm{elu}$ activation.
 	\item A transposed convolutional layer with 9 filters of size (3,1), unit stride, and an $\mathrm{elu}$ activation.
 	\item A transposed convolutional layer with 1 filter of size (3,1), unit stride, and an $\mathrm{elu}$ activation.
\end{enumerate}
\end{decoder}
\begin{decoder}[Used for $1/32$ patches]\label{model:d_incorrect_32}
\begin{enumerate}
    \item A fully connected layer, which is then reshaped to 27 channels, followed by an $\mathrm{elu}$ activation.
    \item An $\mathrm{elu}$ activation layer.
 	\item A transposed convolutional layer with 9 filters of size (3,1), stride of (2,1), and an $\mathrm{elu}$ activation.
 	\item A transposed convolutional layer with 9 filters of size (3,1), unit stride, and an $\mathrm{elu}$ activation.
 	\item A transposed convolutional layer with 1 filter of size (3,1), unit stride, and an $\mathrm{elu}$ activation.
\end{enumerate}
\end{decoder}
\end{document}